\documentclass[aps, pra, twocolumn, tightenlines, letterpaper, amsmath, amssymb, preprintnumbers, floatfix, longbibliography, nofootinbib]{revtex4-2}

\usepackage{dsfont}
\usepackage{amsmath}
\usepackage{amssymb}
\usepackage{physics}
\usepackage{tabu}
\usepackage{tabularx}
\usepackage{bm}
\usepackage{tikz}
\usetikzlibrary{quantikz}
\usepackage{graphicx}
\usepackage{placeins}
\usepackage{textgreek}
\usepackage{multirow}
\usepackage{makecell}
\usepackage{soul}
\usepackage{diagbox}
\usepackage{xcolor}
\usepackage[normalem]{ulem}

\makeatletter
\newcommand{\fmarki}{*}
\newcommand{\fmarkii}{\ensuremath{\dagger}}
\newcommand{\fmarkiii}{\ensuremath{\ddagger}}
\newcommand{\fmarkiv}{\ensuremath{\mathsection}}
\newcommand{\fmarkv}{\ensuremath{\mathparagraph}}
\newcommand{\fmarkvi}{\ensuremath{\|}}

\renewcommand\onecolumngrid{% <<<<<<
\do@columngrid{one}{\@ne}%
\def\set@footnotewidth{\onecolumngrid}% <<<<<<<<<<<<<<<<
\def\footnoterule{\kern-6pt\hrule width 1.5in\kern6pt}%
}

\renewcommand\twocolumngrid{% <<<<<<
        \def\footnoterule{% restore rule
        \dimen@\skip\footins\divide\dimen@\thr@@
        \kern-\dimen@\hrule width.5in\kern\dimen@}
        \do@columngrid{mlt}{\tw@}
}%

\def\si{{}^1\kern-.14em S_0}
\def\siii{{}^3\kern-.14em S_1}
\def\piii{{}^3\kern-.14em P_1}
\def\diii{{}^3\kern-.14em D_1}

\newcommand{\vbar}{\bar{v}}

\def\@fnsymbol#1{{\ifcase#1\or \fmarki\or \fmarkii\or \fmarkiii\or \fmarkiv\or \fmarkv\or \fmarkvi \else\@ctrerr\fi}}
\makeatother

\renewcommand{\fmarkvi}{\$}

\usepackage[hypertexnames=false]{hyperref}
\hypersetup{
    colorlinks=true,       % false: boxed links; true: colored links
    linkcolor=blue,          % color of internal links
    citecolor=blue,        % color of links to bibliography
    filecolor=blue,      % color of file links
    urlcolor=blue           % color of external links
}

\newcolumntype{Y}{>{\centering\arraybackslash}X}

\makeatletter
\pretocmd\frontmatter@thefootnote{\color{black}}{}{}
\makeatother

\usepackage{orcidlink}

\begin{document}

%\title{Stabilizer-Accelerated Ground-State Estimation of Quantum Many-Body Systems}

\title{Stabilizer-Accelerated Quantum Many-Body Ground-State Estimation }

\author{Caroline E.~P.~Robin\,\orcidlink{0000-0001-5487-270X}}
\email{crobin@physik.uni-bielefeld.de}
\affiliation{Fakult\"at f\"ur Physik, Universit\"at Bielefeld, D-33615, Bielefeld, Germany}
\affiliation{GSI Helmholtzzentrum f\"ur Schwerionenforschung, Planckstra{\ss}e 1, 64291 Darmstadt, Germany}

\begin{abstract}

We investigate how the stabilizer formalism, in particular highly-entangled stabilizer states, can be used to describe the emergence of many-body shape collectivity from individual constituents, in a symmetry-preserving and classically efficient way.
The method that we adopt is based on determining an optimal separation of the Hamiltonian into a stabilizer component and a residual part inducing non-stabilizerness. The corresponding stabilizer ground state is efficiently prepared using techniques of graph states and stabilizer tableaux.

We demonstrate this technique in context of the Lipkin-Meshkov-Glick model, a fully-connected spin system presenting a second order phase transition from spherical to deformed state.
The resulting stabilizer ground state is found to capture to a large extent both bi-partite and collective multi-partite entanglement features of the exact solution in the region of large deformation. We also explore several methods for injecting non-stabilizerness into the system, including ADAPT-VQE, and imaginary-time evolution (ITE) techniques. Stabilizer ground states are found to accelerate ITE convergence due to a larger overlap with the exact ground state.

While further investigations are required, the present work suggests that collective features may be associated with high but simple large-scale entanglement which can be captured by stabilizer states, while the interplay with single-particle motion may be responsible for inducing non-stabilizerness. 
This study motivates applications of the proposed approach to more realistic quantum many-body systems, whose stabilizer ground states can be used in combinations with powerful classical many-body techniques and/or quantum methods.

\end{abstract}

\maketitle

%%%%%%%%%%%%%%%%%%%%%%%%%%%%%%%%%%
\section{Introduction}
%%%%%%%%%%%%%%%%%%%%%%%%%%%%%%%%%%

Collective behaviors appear in numerous areas of quantum many-body physics~\cite{sewell1986quantum,bruus2004many,zelevinsky2017physics}.
They include, for example, superconductivity and superfluidity encountered in condensed matter and nuclear physics, various forms of quantum magnetism in spins systems and solid state physics, or superradiant phenomena in quantum optics. Collective vibrational phonons occur in various materials and large nuclei.
Finite mesoscopic systems such as atomic nuclei and metallic clusters can also exhibit spontaneous intrinsic shape deformation~\cite{RevModPhys.74.1283} and dynamical rotations.

Describing how these collective phenomena emerge from the interaction between fundamental constituents is a great challenge, in particular because they typically involve a large degree of entanglement and complexity 
distributed over
many degrees of freedom.
The deformation in atomic nuclei, for example, is known to involve a large number of protons and neutrons behaving in a coherent manner~\cite{PhysRevLett.60.2254,WERNER19941,zelevinsky2017physics}.
This was corroborated by a recent study that found clear correlations between multi-proton-neutron entanglement, in the form of $n$-tangles, and the onset of shape collectivity in nuclei~\cite{Brokemeier:2024lhq}.
Despite impressive progress, these collective features are known to be notoriously difficult to capture within classical {\it ab-initio} methods based on single-particle degrees of freedom, such as configuration-interaction methods~\cite{Navratil:2009ut,BARRETT2013131,schaefer2013methods}, as they require large model spaces and untractable numbers of many-body configurations. 

With advances in the area of quantum computation, it has become clear that quantum computers, coupled with powerful classical devices, provide the most promising ecosystems for developing accurate and precise descriptions of quantum many-body structure and dynamics. As quantum devices can  
embody the quantum complexity of the system of interest in a natural and efficient manner, they constitute ideal tools for describing and studying the emergence of collectivity.

In practice, however, finding strategies to prepare ground states of largely-collective systems on digital quantum computers is not an obvious task~\footnote{It has been shown that finding the ground state of a {\it general} $k$-local ($k$-qubit) Hamiltonian with $k \geq 2$ is QMA-complete~\cite{Kempe:2004sak}. }. While techniques such as the Adaptive Derivative-Assembled Pseudo-Trotter ansatz Variational Quantum Eigensolver (ADAPT-VQE)~\cite{Grimsley:2018wnd} involving individual few-particle excitation operators have been successful in describing quantum chemistry systems, which typically display small degrees of collectivity, they have been found to require numerous iterations to build coherence in collective systems, and, in such cases, are often difficult to converge to a desired accuracy~\cite{Romero:2022blx,Perez-Obiol:2023vod}.

Traditionally, one well-known strategy to describe collective effects in computationally more tractable ways is to break explicit symmetries, which have to be subsequently restored.
Such techniques have been extensively employed in classical calculations, where, for example, particle-number or spherical symmetries are broken to describe superconducting/superfluid pairing and deformation effects~\cite{ring2004nuclear,Yao:2022gsc}.
More recently these techniques have been explored in the context of quantum computing in {\it e.g.} Refs.~\cite{Lacroix:2022vmg,RuizGuzman:2022enk,RuizGuzman:2023sds,LaBorde:2024dle}.
In Ref.~\cite{Robin:2023pgi} we developed an Hamiltonian-Learning Variational Quantum Eigensolver (HL-VQE) algorithm where a symmetry-breaking transformation of the Hamiltonian was used to reduce entanglement and computational complexity of the system. 
In general, the drawback of such strategy is that the symmetry restorations, in particular those related to deformation, are numerically highly costly for large realistic systems. 
\\

At the same time, new directions in the development of many-body methods based on quantum complexity have been developing. This is enabled by progress in the field of quantum information, and the development of measures and techniques for characterizing and quantifying various aspects of complexity in quantum systems.

Entanglement is a long known aspect of quantum complexity, which characterizes how intricately intertwined 
%connected
the sub-components of a system are within the Hilbert space. Systems with no or low entanglement are close to classical, and thus, many-body methods utilizing low-entanglement for efficient classical computations have been advanced~\footnote{Note that various traditional many-body methods have been recently re-interpreted from an entanglement point of view. For example, the Hartree-Fock (HF) or mean-field technique consists in unitarily transforming the single-particle basis, to find an approximate energy-minimizing solution which is unentangled in this new basis. More generally, connections between basis changes and entanglement minimization have been made, see {\it e.g.} Refs.~\cite{Szalay_2015,PhysRevA.92.042326,Robin:2020aeh,Bulgac:2022cjg}.}, the most successful example of which is tensor networks~\cite{Banuls2023}.
Such methods, originally designed for low-dimension many-body systems, however become prohibitively expensive 
in systems with strong and/or collective entanglement.

On the other hand, it has been realized in the last decades 
that entanglement alone does not provide a complete characterization of quantum complexity, as some quantum states, known as stabilizer states, can exhibit maximal large-scale entanglement, while being simple in nature and efficiently preparable on a classical computer~\cite{gottesman1998heisenberg,Aaronson_2004}.
What cannot be captured classically is the interplay of entanglement with non-stabilizerness (also commonly known as "magic"~\footnote{In the present manuscript, we use the terms non-stabilizerness and magic interchangeably.}~\cite{Bravyi_2005}).
In order to fully specify the quantum complexity of a state, and assess the need for quantum computers, both characterizations of entanglement and non-stabilizerness are therefore required.

The recent development of magic measures, in particular those based on R\'enyi entropies~\cite{Leone:2021rzd}, have allowed for investigations of non-stabilizerness and connections to physical phenomena in various quantum many-body systems. 
Understanding how quantum complexity evolves during the transition from single-particle to collective regimes is particularly important for addressing the question of collectivity emergence.
In this context, the magic of 
a variety of spin models~\cite{Oliviero:2022euv,Haug:2022vpg,Rattacaso:2023kzm, Haug:2023hcs,Tarabunga:2023ggd,Tarabunga:2023xmv,Passarelli:2024tyi,Frau:2024qmf,Catalano:2024bdh,Viscardi:2025vya} and gauge theories~\cite{Tarabunga:2023ggd,Falcao:2024msg,Hartse:2024qrv} 
have been studied in relation with phase transitions and thermalization.
In nuclear physics, the magic in ground states of atomic nuclei was investigated in Ref.~\cite{Brokemeier:2024lhq} in connection with entanglement and collectivity, while the magic power of nuclear and hypernuclear forces was studied in Ref.~\cite{Robin:2024bdz}.
Recently, non-stabilizerness was shown to be connected with molecular bonding in Refs.~\cite{Gu:2023ylw,Sarkis:2025oab}.
Magic evolution during many-body three-flavour neutrino propagation was studied in Ref.~\cite{Chernyshev:2024pqy}.
In high-energy particle physics, Ref.~\cite{White:2024nuc} studied the magic in the production of top quarks at the LHC. 
Magic was also investigated within quantum gravity~\cite{Cepollaro:2024qln,Cao:2024nrx}, and in QED particle-scattering~\cite{Liu:2025qfl}.

Further,
the stabilizer formalism, initially introduced in the context of quantum error correction~\cite{gottesman1998heisenberg,Aaronson_2004},
is conceivably helpful in the development of quantum many-body simulations. 
In the context of collective systems, in particular, the large-scale entanglement exhibited by stabilizer states may provide a way to capture collective features of the system in a computationally tractable way, and thus potentially provide optimal starting points to classical and/or quantum computations.

The idea of leveraging the stabilizer formalism for describing quantum many-body systems is so far rather under-explored, although rapidly developing.
Methods utilizing entanglement-magic separation to augment classical tensor network computations 
include the stabilizer tensor networks~\cite{PhysRevLett.125.241602,PhysRevB.105.104306,Masot-Llima:2024doz,Nakhl:2024gfr}, 
as well as Clifford-augmented matrix product states~\cite{PRXQuantum.6.010345,PhysRevLett.133.150604,Qian:2024vea,PhysRevLett.134.150403,PhysRevLett.134.150404,Huang:2024ron} which were applied to spin systems and Hubbard model.
Techniques in which stabilizer states are used as initial states further refined by subsequent quantum circuits have been developed in the context of one-dimensional local systems in Ref.~\cite{Sun:2024lts}.
In quantum chemistry, quantum algorithms involving entanglement-magic separation have been developed via injections of magic on top of stabilizer states, see {\it e.g.} Refs.~\cite{10.1145/3567955.3567958,Bhattacharyya:2023xad,Gu:2023ylw}, 
or via Clifford transformations of the Hamiltonian, see {\it e.g.} Refs.~\cite{Mishmash:2023eow,Anand:2023ofu,Gu:2024ure}. 
Non-contextual VQE based on projections onto stabilizer subpaces has also been developed and applied to a range of small molecules~\cite{Kirby2021contextualsubspace,Weaving:2022hbz}.
The systems investigated in the aforementioned studies, however, are expected to display local entanglement features and/or a small degree of collectivity. For example, quantum chemistry systems typically exhibit rather pure wave functions, in comparison to the large collectivity observed in nuclei which brings into play a much larger number of configurations, notably due to the interplay between the two fermion species (protons and neutrons). \\

The goal of the present work is to start investigating whether the stabilizer formalism provides a way to describe the emergence of collectivity in an efficient manner.
In particular we explore how stabilizer states can capture collective deformation effects, while preserving the symmetries of the system explicitly, and thus avoiding the need for projection techniques.

For demonstration we employ the Lipkin-Meshkov-Glick (LMG) model, a fully-connected spin system which displays a phase transition analogous to the transition to deformed nuclei, and thus constitutes an ideal test case for this study.
This model is also 
employed in condensed matter physics to describe two-mode Bose-Einstein condensates~\cite{PEREZCAMPOS2010325,Chen:09,PhysRevA.91.053612}, for the production of spin-squeezed states relevant to quantum metrology and sensing~\cite{PhysRevA.90.022111,PhysRevA.109.052618}, and is also relevant to trapped-ion quantum computing due to its all-to-all connectivity.

The method that we adopt, which we describe in Sec.~\ref{sec:general}, consists in determining an optimal stabilizer Hamiltonian by identifying energy-minimizing groups of mutually-commuting operators. The corresponding stabilizer ground state is then classically efficiently prepared using techniques of graph states and stabilizer tableaux.  
In Sec.~\ref{sec:Stabilizer_gs_LMG}, we apply this method explicitly to the LMG model, and explore to what extent the stabilizer ground state can capture properties of the exact solution across the phase transition, including bi-partite and collective multi-partite entanglement.
In Sec.~\ref{sec:Beyond_stabilizer}, we explore a few techniques to inject magic on top of the stabilizer ground state, including ADAPT-VQE, and imaginary-time evolution.
Finally, Sec.~\ref{sec:conclu} provides a summary and perspectives to this work.

%%%%%%%%%%%%%%%%%%%%%%%%%%%%%%%%%%
\section{Method}
\label{sec:general}
%%%%%%%%%%%%%%%%%%%%%%%%%%%%%%%%%%

%
The strategy adopted in the present work consists in searching for an optimal division of the Hamiltonian into a stabilizer part $H_{stab}$ plus a part $W$ inducing non-stabilizerness, so that the ground state of $H_{stab}$, which can be efficiently prepared classically, will ideally provide a close approximation to the exact ground state. 
To be treated exactly, the non-stabilizer term $W$ will, in principle, require amounts of classical resources which scale exponentially with system size or required precision. Such term can be treated subsequently (in part), either using refined 
classical algorithms, or, ultimately, on the quantum device using an adequate quantum or classical-quantum algorithm of choice. Such strategy based on Hamiltonian division has been employed in different contexts in, {\it e.g.}, Refs.~\cite{Kirby2021contextualsubspace,Weaving:2022hbz,Bhattacharyya:2023xad,Gu:2024ure,Sun:2024lts}.

Let us consider a general many-body Hamiltonian, mapped onto $N$-qubit strings of Pauli operators $\hat P$:
\begin{equation}
        \hat H = \sum_{P \in \mathcal{G}_N(H)} a_P \hat P \; ,
\end{equation}
where $\mathcal{G}_N(H) \subset \mathcal{G}_N$ is the set of Pauli strings that map $\hat H$, 
which is a subset of the generalized Pauli group $\mathcal{G}_N$
\begin{equation}
    \mathcal{G}_N = \{ \varphi \,  \sigma^{(1)} \otimes \sigma^{(2)} ... \otimes \sigma^{(N)}  \} \; ,
\end{equation}
where $\sigma^{(j)} \in \{ \mathds{1}, X_j, Y_j, Z_j \}$ is a Pauli operator acting on qubit $j$, and  $\varphi \in \{ \pm 1, \pm i\}$.

We wish to separate the Hamiltonian into
\begin{align}
\hat H 
       &= \underbrace{ \sum_{P \in \mathcal{S}} a_P \hat P }_{\hat{H}_{stab}} +  \underbrace{\sum_{P \in \mathcal{G}_N(H) \notin \mathcal{S}} a_P \hat P}_{\hat W} \; , 
\label{eq:Hami_sep_gen}
\end{align}
where $\mathcal{S}$ forms a stabilizer group with $2^N$ commuting elements~\footnote{Possibly with some of the coefficients $a_P$ equal to zero to form a complete group with $2^N$ elements.}.

The ground state $\ket{\Psi_s}$ of $\hat{H}_{stab}$ is stabilized by the $2^N$ operators in $\mathcal{S}$ and thus is, by definition, a stabilizer state which can be exactly specified by the $N$ generators of that group. 
This latter property makes it efficient to prepare $\ket{\Psi_s}$ with a classical computer using the stabilizer tableau formalism~\cite{gottesman1998heisenberg,PhysRevA.57.127,Aaronson_2004}.

The decomposition in Eq.~\eqref{eq:Hami_sep_gen} is not unique, as there are several commuting groups in  $\mathcal{G}_N(H)$ that one can pick to form $\mathcal{S}$. Ideally, as mentioned above, we want the ground state of $\hat{H}_{stab}$ to optimally approximate the exact ground state of the full Hamiltonian. Here we select an energy criterion to achieve this task, {\it i.e.} we choose $\mathcal{S}$ so that $\ket{\Psi_s}$ be the stabilizer state minimizing the energy of the full Hamiltonian $\langle \Psi_s |\hat H| \Psi_s \rangle$ ~\footnote{Of course, it may be that such energy criterion does not give the best approximation to the wave function, or best starting point to include $W$, and one may want to explore other criteria. These are however typically more difficult to implement without knowledge of the final answer and we do not attempt it here.}.

It is useful to note that, due to the specific properties of stabilizer states, the stabilizer ground state energy can be determined from the stabilizer group $\mathcal{S}$, without knowledge of the state itself.
Indeed, it is known that, for a stabilizer state $\ket{\Psi_s}$, the expectation value of a Pauli string $\langle \Psi_s |\hat P| \Psi_s \rangle $ can only take values $\pm 1$ (if $\hat P$ stabilizes or "anti-stabilizes $\ket{\Psi_s}$") or $0$~\cite{zhu2016clifford}:
\begin{align}
    \langle \Psi_s |\hat P| \Psi_s \rangle =
    \begin{cases}
      \pm 1 & \text{if } \hat P \ket{\Psi_s} = \pm \ket{\Psi_s} \; ,\\
      0 & \text{otherwise} \; .\\
    \end{cases}
\end{align}
Thus, the energy of a stabilizer state is simply given by
\begin{equation}
        E\left( \Psi_s  \right) = \langle \Psi_s |\hat H| \Psi_s \rangle = \sum_{P\in \mathcal{G}_N(H)} a_P \underbrace{\langle \Psi_s |\hat P| \Psi_s \rangle}_{0,  \pm 1} \; .
\end{equation}
Since $\ket{\Psi_s}$ is uniquely defined by its stabilizer group $\mathcal{S}$, this equivalently defines the energy of the stabilizer group $\mathcal S$~\cite{Sun:2024lts}:
\begin{align}
        E(\mathcal{S}) &= \sum_{P \in \mathcal{G}_N(H)} a_P \, E(\hat P, \mathcal{S}) \; , \nonumber \\
        & \text{where } 
        E(\hat P, \mathcal{S}) = 
        \begin{cases} 
        \pm 1 & \text{if } \pm \hat P \in  \mathcal{S} \; , \\
         0 & \text{ otherwise}  \nonumber \; .
        \end{cases}
\end{align}

In summary, the stabilizer Hamiltonian and corresponding stabilizer ground-state energy can be found by partitioning the operators in $\mathcal{G}_N(H)$ into commuting 
subsets $Q_k \subset \mathcal{G}_N(H)$ (excluding $-I$ to ensure $\mathcal{S}$ is a stabilizer group), and evaluating the stabilizer energy of each subset, allowing for all sign possibilities. The set $Q_k^{min}$ with the lowest energy is chosen to form the stabilizer Hamiltonian $\hat{H}_{stab}$ in Eq.~\eqref{eq:Hami_sep_gen}.
The complete stabilizer group $\mathcal{S}$ can be obtained by extracting from $Q_k^{min}$ the maximal number of elements that are linearly independent. These elements, which we denote $g_1, ..., g_m$, constitute $m \leq N$ generators of $\mathcal{S}$. They can be further completed with commuting operators $g_{m+1}, .., g_N$ to generate the complete stabilizer group $S = \langle g_1, ..., g_N \rangle$, and fully specify the corresponding stabilizer ground state $\ket{\Psi_s}$.
\\

In order to prepare the stabilizer ground state, one can build on the knowledge of graph states, and make use of the fact that any (entangled) stabilizer state $\ket{\Psi_s}$ is Clifford-locally equivalent to a graph state~\cite{10.5555/2011477.2011481,Grassl}.
In general, an $N$-qubit graph state $\ket{G}$ is associated with a simple graph $G = (V,E)$, where $V$ denotes the sets of vertices ($N$ qubits) and $E$ denotes the set of edges between these vertices.
The graph state $\ket{G}$ can be prepared as
\begin{align}
    \ket{G} = \left( \prod_{e \in E } \text{CZ}_e \right) \ket{+}^{\otimes N} \; ,
    \label{eq:Gdef}
\end{align}
and is stabilized by operators of the form~\cite{ed1d1be3cb1a475fb9d87d2369dfef15}
\begin{align}
    g_i^{G} = X_i \prod_{j \in n_i} Z_j \; .
    \label{eq:stabG_def}
\end{align}
In Eq.~\eqref{eq:Gdef} $\text{CZ}_e$ denotes a controlled-$Z$ gate acting between the two qubits connected by edge $e$ and $\ket{+} = (\ket{0} + \ket{1})/\sqrt{2}$, while in Eq.~\eqref{eq:stabG_def} $n_i$ denotes the set of vertices (qubits) connected to vertex $i$ by an edge.
A stabilizer state $\ket{\Psi_s}$ can be implemented by acting on a graph state $\ket{G}$ with local Clifford unitaries, {\it i.e.}
\begin{equation}
    \ket{\Psi_s} = \prod_{i=1}^N C_i \ket{G} = \prod_{i=1}^N C_i \,  \prod_{e \in E} \text{CZ}_e \ket{+}^{\otimes N} \; ,
\end{equation}
where the $C_i$ are single-qubit Clifford operators~\cite{10.5555/2011477.2011481,Grassl}.

In a general case, one can, for example, treat the single-qubit Clifford operations $C_i$ as (discrete) variational parameters and optimize according to an energy minimization. Such optimization procedures, involving discrete parameters, can, however, be difficult to converge.
In this work, we will instead employ an efficient procedure for determining the $C_i$'s based on the stabilizer tableau formalism~\cite{PhysRevA.69.022316}.
In the case of the LMG model employed below, we will see that one can also use intuition to conjecture 
a preparation for $\ket{\Psi_s}$.
In any case, once an expression of the stabilizer state in terms of Clifford operators acting on $\ket{0}^{\otimes N}$ has been found, this state can be efficiently prepared  with a classical computer.
\\

Finally as a last step, one can treat the term $\hat W = \sum_{P \in \mathcal{G}_N(H) \notin \mathcal{S}} a_P \hat P$ in Eq.~\eqref{eq:Hami_sep_gen} via a method of choice, that is to be chosen based on the physical properties of the system.

%%%%%%%%%%%%%%%%%%%%%%%%%%%%%%%%%%%%%%%%%%%%%%%%%%%%%%%%%%%%%%%%%%%%
%%%%%%%%%%%%%%%%%%%%%%%%%%%%%%%%%%%%%%%%%%%%%%%%%%%%%%%%%%%%%%%%%%%%
%%%%%%%%%%%%%%%%%%%%%%%%%%%%%%%%%%%%%%%%%%%%%%%%%%%%%%%%%%%%%%%%%%%%
\section{Stabilizer Ground States in the Lipkin-Meshkov-Glick Model}
\label{sec:Stabilizer_gs_LMG}
%%%%%%%%%%%%%%%%%%%%%%%%%%%%%%%%%%%%%%%%%%%%%%%%%%%%%%%%%%%%%%%%%%%%
%%%%%%%%%%%%%%%%%%%%%%%%%%%%%%%%%%%%%%%%%%%%%%%%%%%%%%%%%%%%%%%%%%%%
%%%%%%%%%%%%%%%%%%%%%%%%%%%%%%%%%%%%%%%%%%%%%%%%%%%%%%%%%%%%%%%%%%%%

The LMG model~\cite{LIPKIN1965188} originally described a system of $N$ identical fermions distributed on two $N$-fold degenerate shells separated by an energy gap $\varepsilon$. Alternatively, this system can be mapped onto a system of $N$ spins in an external field along the $z$ direction, and interacting in the $xy$ plane with all to all connectivity. The Hamiltonian can then be written as~\cite{Hengstenberg:2023ryt}
\begin{eqnarray}
\hat H 
= \varepsilon \hat{J}_z 
    - V_x (\hat J_x^2 + \chi \hat J_y^2) 
\; , 
\label{eq:LMG_Hami}
\end{eqnarray}
where the collective spin operators are given by
\begin{equation}
    \hat J_\alpha = \frac{1}{2} \sum_{i=1}^N \hat \sigma_\alpha^{(i)} \; , \hspace{1cm} \alpha = x,y,z \; ,
\end{equation}
where $\sigma_\alpha^{(i)}$ denotes a Pauli operator acting on spin $i$.
The Hamiltonian in Eq.~\eqref{eq:LMG_Hami} preserves a number of symmetries. In particular, since the interaction flips spins by pairs, it preserves the parity of the number of spins pointing up (in the direction of the external field). This symmetry is associated with operator 
\begin{equation}
    \hat{\Pi} = e^{i \pi \hat{N}_+} \sim \prod_i Z_i \; , 
    \label{eq:parity_op} 
\end{equation}
where $\hat{N}_+ = \hat{J_z} + \hat{N}/2$ counts the number of spins up.
\\
In the following, we will work with the dimensionless Hamiltonian $\widetilde{H} = \hat{H} / \varepsilon$ and rescaled interaction strength $\vbar_x = \frac{(N-1)V_x}{\varepsilon}$. \\

We will focus on the case $\vbar_x > 0$, which corresponds to a ferromagnetic coupling~\cite{PhysRevA.69.022107}. 
In the mean-field limit, the system undergoes a second-order phase transition at $\vbar_x=1$ between a normal phase ($\vbar_x<1$, single-particle regime) and a phase where parity symmetry is broken ($\vbar_x>1$, collective regime). This phase transition is analogous to the transition from a spherical to a deformed nucleus.

In the case  $\chi \in [-1, 0)$, in particular for $\chi = -1$ most studied in nuclear physics, the interacting ground state wave function typically expands over several collective Dicke (angular momentum) states $\ket{J=N/2, J_z}$. 
In the limit of large $N$, in the largely-deformed phase, the energy of the system is known to be well described by a deformed Hartree-Fock state (coherent SU(2) state).
The case $\chi = 0$ leading to spin squeezed states is most relevant for quantum sensing and metrology (see {\it e.g.} \cite{PhysRevA.90.022111,PhysRevA.109.052618}).
In the isotropic case $\chi = 1$, the exact ground state reduces to a single Dicke state with $J_z$ value determined by the interaction strength.
In the following, we will focus on the parameter region $\chi \in [-1, 0)$ which is most relevant to simulations of realistic many-body systems.
%
%

%%%%%%%%%%%%%%%%%%%%%%%%%%%%%%%%%%%%%%%%%%%%%%%%%%%%%%%
\subsection{Stabilizer Hamiltonian} 
%%%%%%%%%%%%%%%%%%%%%%%%%%%%%%%%%%%%%%%%%%%%%%%%%%%%%%%
The procedure to find the stabilizer Hamiltonian, is described in details in appendix~\ref{sec:app_stabil_Hami}, starting from $N=2$ and increasing system size. 
Below we summarize the results. 
\\

The first step consists in mapping the Hamiltonian in Eq.~\eqref{eq:LMG_Hami} to qubits.
Using a direct spin-to-qubit mapping, we obtain 
\begin{align}
    \widetilde{H} 
    = \frac{1}{2} \sum_i Z_i -  \frac{\vbar_x}{2(N-1)} \sum_{i<j}( X_i X_j + \chi \, Y_i Y_j) \; ,
    \label{eq:Hami_chi}
\end{align}
where we now use the usual notations $X_i$, $Y_i$, $Z_i$ to denote the Pauli operators acting on qubit $i$.
\\

For arbitrary $N>2$, we can extract several subsets $Q_i^{(N)}$ of mutually-commuting operators from the Hamiltonian in Eq.~\eqref{eq:Hami_chi}. Each of these $Q_i^{(N)}$ can be augmented by commuting operators (excluding $-\mathds{1}$) in order to form a stabilizer group $\mathcal{S}$ with dimension $2^N$. In particular, limiting ourselves to subsets that satisfy permutation symmetry, we find three relevant subsets, which are listed below.

\begin{enumerate}

    \item $Q_1^{(N)} = \{ -Z_1, -Z_2, ... , -Z_N \}$, corresponding to $H_{stab} = \frac{1}{2} \sum_i Z_i$ with stabilizer energy $E_{s,1}(N) = -N/2$. 
    The $N$ operators in $Q_1^{(N)}$ generate the stabilizer group $\mathcal{S}_1^{(N)} = \langle -Z_1, -Z_2, ... , -Z_N \rangle$ which stabilizes the non interacting ground state $\ket{\Psi_{s,1}}^{(N)} = \ket{\downarrow \downarrow ... \downarrow} \equiv \ket{11...1} \equiv \ket{1}^{\otimes N}$.
    
    \item \label{case:2} $Q_2^{(N)} = \{ X_1 X_2, X_1 X_3, ..., X_2 X_3 ... , X_{N-1} X_N  \}$, corresponding to $H_{stab} = -  \frac{\vbar_x}{2(N-1)} \sum_{i<j} X_i X_j$, with stabilizer energy $E_{s,2}(N) = - \frac{N\bar{v}_x}{4}$. 
    Among the $N(N-1)/2$ operators in $Q_2^{(N)}$, only $N-1$ of them are linearly independent.
    A stabilizer group $\mathcal{S}_2^{(N)}$ can then be formed by complementing these chosen $N-1$ independent operators with a $N$-th commuting operator, such as $(-1)^N \, Z_1 Z_2... Z_N$, which satisfies the desired permutation and parity symmetries.
    We choose $\mathcal{S}_2^{(N)} = \langle X_1 X_N, X_2 X_N, ..., X_{N-1}X_N, (-1)^N \, Z_1 Z_2... Z_N \rangle $.
    For $N$ even (resp. odd), the state stabilized by this group is an entangled state $\ket{\Psi_{s,2}}^{(N)}$ corresponding to an equal superposition of all computational-basis states with even (resp. odd) number of spins down (1's).
    For example, for $N=3$, 
    \begin{align}
    \ket{\Psi_{s,2}}^{(3)} =  \frac{1}{2} (\ket{111} + \ket{100} + \ket{010} + \ket{001}) \; ,
    \label{eq:stab_N3}
    \end{align}
    while for $N=4$,
    \begin{align}
    \ket{\Psi_{s,2}}^{(4)} = \frac{1}{2 \sqrt{2}} &\Bigl( \ket{0000} + \ket{0011} + \ket{0110} + \ket{1100}  \nonumber \\ 
    & + \ket{0101} + \ket{1111} + \ket{1010} + \ket{1001} \Bigr)\; .
    \label{eq:stab_N4}
    \end{align}
    
    \item $Q_3^{(N)} = \{ Y_1 Y_2, Y_1 Y_3, ..., Y_2 Y_3 ... , Y_{N-1} Y_N  \}$, corresponding to $H_{stab} =   - \frac{\vbar_x \chi}{2(N-1)} \sum_{i<j} Y_i Y_j$, with stabilizer energy $E_{s,3}(N) = -\frac{N\vbar_x \chi}{4}$. 
    Similarly to case~\ref{case:2}. above, one can form a stabilizer group $\mathcal{S}_3^{(N)} = \langle Y_1 Y_N, Y_2 Y_N, ..., Y_{N-1} Y_N, (-1)^N \, Z_1 Z_2... Z_N \rangle $.
    The corresponding stabilizer states $\ket{\Psi_{s,3}}^{(N)}$ have a similar form as $\ket{\Psi_{s,2}}^{(N)}$, with different relative signs between the components.
    Depending on the value of $\chi$ and $N$ one can change the signs in front of the $Y_iY_j$ terms to bring the stabilizer energy $E_{s,3}(N)$  down, however, in general, this choice of subset will yield a larger energy than $Q_2^{(N)}$ (except in the case $\chi=+1$ for which they are degenerate). 
\end{enumerate}

In summary, for a system with $N>2$ spins, if $-\frac{N \vbar_x}{4} > -\frac{N}{2}$, i.e., $\vbar_x < 2 $, the lowest-energy stabilizer state is $\ket{\Psi_{s,1}}^{(N)} = \ket{1}^{\otimes N}$ with stabilizer energy $E_{s,1}(N)= -N/2$, while if $\vbar_x > 2$ the lowest-energy stabilizer state is an entangled state $\ket{\Psi_{s,2}}^{(N)}$ with energy $E_{s,2}(N)=-\frac{N \vbar}{4}$. 
\\

We note that there are other commuting subsets of operators that we did not consider here, such as e.g. $\{ X_1 X_2, -Y_1 Y_2, -Z_3, -Z_4...-Z_N \}$. Such grouping would break permutation invariance, which is not desired, and the corresponding stabilizer energy would be $-\frac{(N-2)}{2}-\frac{\vbar_x (1-\chi)}{2(N-1)} $ which is always greater than $E_{s,1}(N)$ and/or $E_{s,2}(N)$ above.
\\

The case $N=2$ is special and presents a transition from unentangled to entangled stabilizer state occurring at $\vbar_x =1$ (see appendix~\ref{sec:app_stabil_Hami}.)
\\

%%%%%%%%%%%%%%%%%%%%%%%%%%%%%%%%%%%%%%%%%%%%%%%%%%%%%%%
\subsection{Stabilizer Ground-State Preparation}
%%%%%%%%%%%%%%%%%%%%%%%%%%%%%%%%%%%%%%%%%%%%%%%%%%%%%%%
The procedure described above provides a decomposition of the Hamiltonian into a stabilizer part $H_{stab}$ plus a magic perturbation $W$, as well as the corresponding stabilizer ground state energy.
While the preparation of the unentangled stabilizer state $\ket{\Psi_{s,1}}^{(N)} = \ket{1}^{\otimes N}$ is trivial, preparing the entangled stabilizer state $\ket{\Psi_{s,2}}^{(N)}$ is less obvious. 
\\

As described in appendix~\ref{sec:app_stabil_Hami}, in the case of the LMG model, one can infer from intuition a way of preparing $\ket{\Psi_{s,2}}^{(N)}$ efficiently classically, using Clifford operators CX, H and X acting on the unentangled state $\ket{0}^{\otimes N}$ as
\begin{align}
        \ket{\Psi_{s,2}}^{(N)} = 
        \begin{cases} 
        \prod_{i<j} \text{CX}_{ji} \ket{0} \otimes \ket{+}^{\otimes (N-1)} & \text{for even } N \; , \\ \\
         \prod_{i<j} \text{CX}_{ji} \ket{1} \otimes \ket{+}^{\otimes (N-1)} & \text{for odd } N  \; .
        \end{cases}
\label{eq:stab_LMG} 
\end{align}
It is easy to see that Eq.~\eqref{eq:stab_LMG} applied to $N=3$ and $N=4$ indeed prepares the states in Eq.~\eqref{eq:stab_N3} and \eqref{eq:stab_N4}, respectively.
\\

\begin{figure}[h]
     \centering
         \includegraphics[width=0.65\columnwidth]{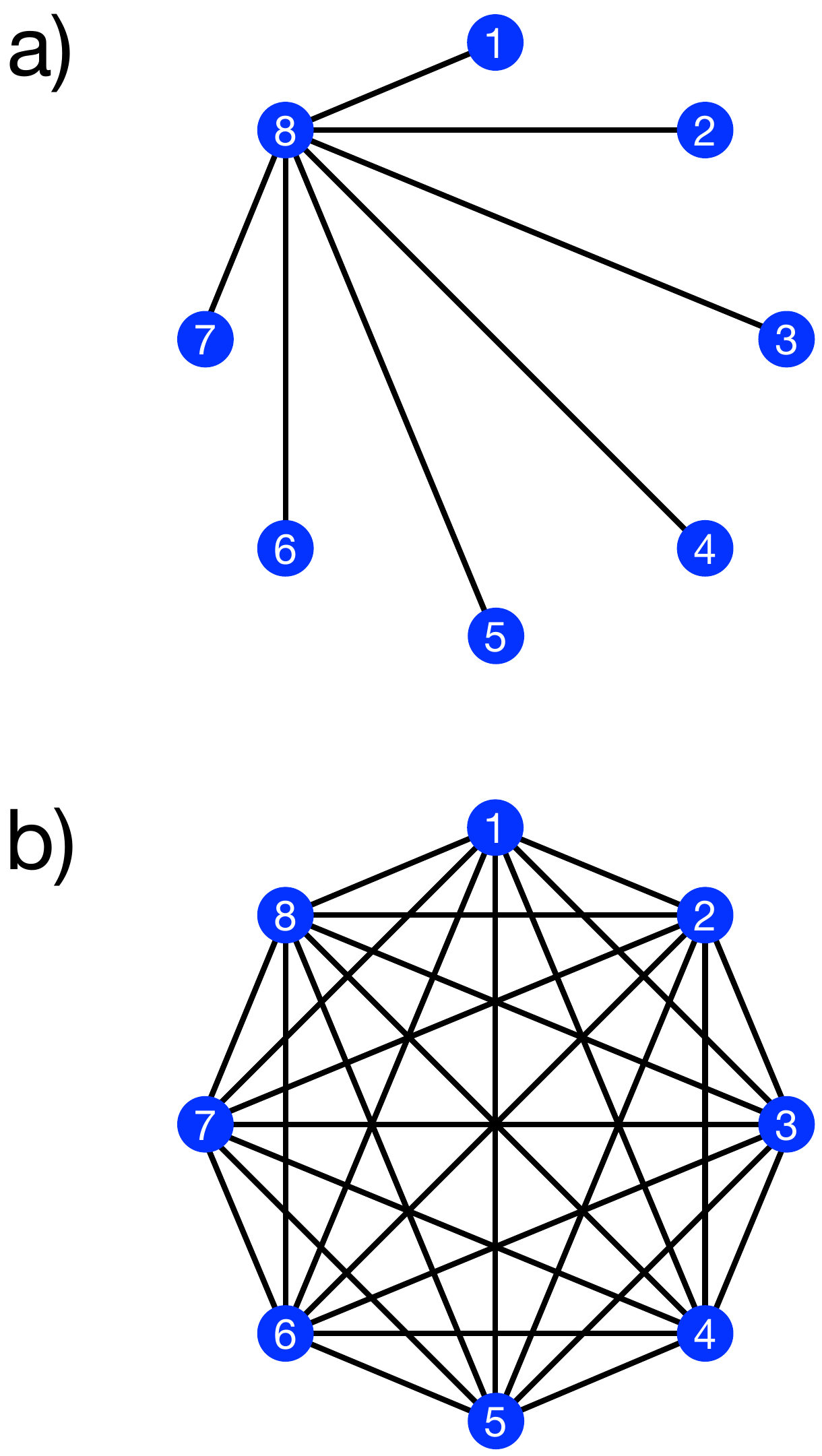}
         \caption{Panel a): graph $G$ associated with the state in Eq.~\eqref{eq:graph_LMG} for $N=8$. Panel b): complementary graph $G^c$ (see more details in appendix~\ref{sec:app_stabil_Hami}). }
         \label{fig:graph_N8}
\end{figure}

More rigorously, it is known that any stabilizer state can be related to a graph state via local Clifford operations~\cite{10.5555/2011477.2011481,Grassl}. Moreover, the corresponding Clifford unitary can be calculated efficiently, and a procedure is described in Ref.~\cite{PhysRevA.69.022316}. This procedure makes use of the stabilizer tableau formalism and provides a systematic way to express the stabilizer ground state in terms of a graph state.
Applying this method to the present LMG model (see details in appendix~\ref{sec:app_stabil_Hami}), we find that 
\begin{align}
    \ket{\Psi_{s,2}}^{(N)} = (X_1)^{N \, \text{mod}\, 2} \, H_N \ket{G} \; ,
    \label{eq:stab_LMG_graph}
\end{align}
where $\ket{G}$ is the graph state given by
\begin{align}
\ket{G} = \prod_{i=1}^{N-1} \text{CZ}_{i, N} \ket{+}^{\otimes N} \; ,
\label{eq:graph_LMG}
\end{align}
which is associated with the graph with edges between qubit $N$ and all qubits $i<N$, as shown in panel a) of Fig.~\ref{fig:graph_N8}.
\\

\noindent $\ket{G}$ is the state stabilized by $\mathcal{S}_G^{(N)}=\langle X_1 Z_N, X_2 Z_N, ..., X_{N-1}Z_N, Z_1 Z_2... X_N \rangle$ which is obtained by conjugating 
$\langle X_1 X_N, X_2 X_N, ..., X_{N-1}X_N, Z_1 Z_2... Z_N \rangle$ with $H_N$.
The gate $X_1$ is applied for odd values of $N$ to provide the negative sign of the last generator $(-)^N Z_1 Z_2... Z_N$, and obtain the correct parity (odd numbers of spins down).
\\

The procedure described in Ref.~\cite{PhysRevA.69.022316}, which can be applied to any Hamiltonian, provides a systematic recipe for preparing the stabilizer ground state from a graph state, avoiding discrete optimization procedures which are often more challenging than continuous ones. Application to more general Hamiltonians will be explored in future works.
\\

%%%%%%%%%%%%%%%%%%%%%%%%%%%%%%%%%%%%%%%%%%%%%%%%%%%%%%%
\subsection{Results}
%%%%%%%%%%%%%%%%%%%%%%%%%%%%%%%%%%%%%%%%%%%%%%%%%%%%%%%

\begin{figure}[h!]
     \centering
         \includegraphics[width=\columnwidth]{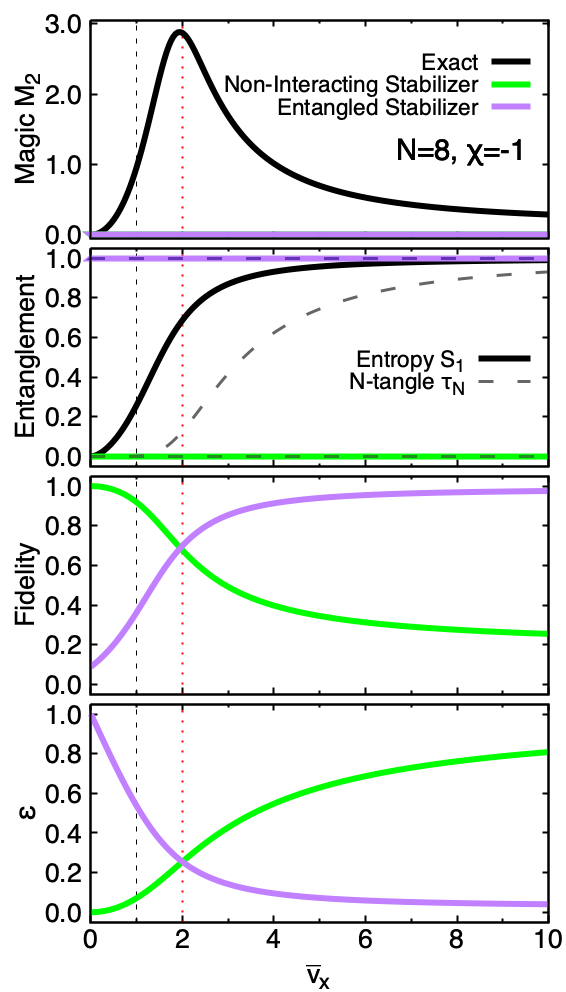}
         \caption{From bottom to top panel: relative energy difference $\varepsilon$, fidelity, entanglement, and stabilizer $2$-R\'enyi entropy $\mathcal{M}_2$, in a system with $N=8$ spins, as a function of the interaction strength $\vbar_x$. 
         The exact solution is shown with black curves, the unentangled (non-interacting) stabilizer state $\ket{\Psi_{s,1}}^{(N)} = \ket{1}^{\otimes N}$ is shown with green curves and the entangled stabilizer state $\ket{\Psi_{s,2}}^{(N)}$ is shown with purple curves. 
         In the entanglement panel, the von Neumann entropy is displayed with plain lines while the $N$-tangle $\tau_N$ is shown with dashed lines.
         The black dashed vertical line denotes the critical point between normal (spherical) and parity-broken (deformed) phases, while the red dotted vertical line denotes the transition from unentangled to entangled stabilizer ground state.
         }
         \label{fig:Stabil_N8_chi-1}
\end{figure}

As an example, we show in Fig.~\ref{fig:Stabil_N8_chi-1} %and \ref{fig:Stabil_N8_chi-0p1} 
results obtained for a system of $N=8$ spins with parameter value $\chi=-1$. % and $\chi=-0.1$, respectively. 
Systems with sizes from $N=2$ up to $N=30$ are provided in appendix~\ref{sec:app_results_varQITP}. 
In particular, for each of the two stabilizer states $\ket{\Psi_{s,i}}^{(N)}$ ($i=1,2$), we analyze the relative energy difference $\varepsilon =  |(E_{s,i} - E_{ex}) / E_{ex}|$ between stabilizer ground-state energy and exact energy, and the fidelity $|\langle  \Phi_{ex} | \Psi_{s,i}\rangle|$ between the stabilizer states $| \Psi_{s,i}\rangle$ and the exact wave function $| \Phi_{ex}\rangle$. 
We also show in the same figure the magic and entanglement features of the exact and stabilizer wave functions.

The magic is 
quantified via stabilizer R\'enyi entropy (SRE) defined according to Ref.~\cite{Leone:2021rzd} as
\begin{align}
    \mathcal{M}_\alpha(\ket{\Psi}) = -\text{log}_2 \, d + \frac{1}{1-\alpha} \text{log}_2 \, \left( \sum_{P \in \tilde{\mathcal{G}}_n} \Xi_P^\alpha \right) \; ,
\end{align}
where $d= 2^N$, $\tilde{\mathcal{G}}_N \subset \mathcal{G}_N$ is the group of Pauli strings with phases $+1$ and $\Xi_P = \langle \Psi |\hat P | \Psi \rangle^2 /d$.
Specifically, we choose the stabilizer $2$-R\'enyi entropy $\mathcal{M}_2(\ket{\Psi})$ which is known to be related to the distance between $\ket{\Psi}$ and the closest stabilizer state~\cite{Haug:2024ptu}, and has been shown to satisfy the required properties of good measures, including monoticity~\cite{Haug:2023hcs,Leone:2024lfr}. 

Entanglement is quantified via (1-spin) von Neumann entropy defined as 
\begin{equation}
    S_1^{(N)} = - \mbox{Tr} \left( \rho_1^{(N)} \text{log}_2 \, \rho_1^{(N)} \right) \; ,
\end{equation}
where $\rho_1^{(N)}$ is the one-spin reduced density matrix that takes the expression~\cite{Hengstenberg:2023ryt}
\begin{eqnarray}
\rho_1^{(N)}= 
\begin{pmatrix}
1-\langle \hat {N}_+ \rangle /N  & 0 \\
0  & \langle \hat N_+\rangle  /N 
\end{pmatrix}
\; ,
\label{eq:1sp_RDM}
\end{eqnarray}
where $\hat N_+ =   N/2 + \hat J_z$ counts the number of spins up.
To quantify multipartite collective entanglement we choose the $n$-tangles~\cite{PhysRevA.61.052306,PhysRevA.63.044301}. For a permutation invariant system, they are simply given by
\begin{align}
    \tau_n = | \langle \Psi | Y^{\otimes n}| \Psi^* \rangle |^2 \; .
    \label{eq:tangles}
\end{align}

We note from Fig.~\ref{fig:Stabil_N8_chi-1} that the entangled stabilizer state $\ket{\Psi_{s,2}}^{(N)}$  
is maximally entangled, in both the bi-partite and $N$-partite sense,
for the whole range of interaction strength. 
By definition, both stabilizer states display constant zero magic.

In the regime below the phase transition $\vbar_x < 1$, the system is dominated by single-particle dynamics and is in the spherical (normal) phase. In this region, the exact state exhibits low entanglement and magic, and is best described by the non-interacting (unentangled) stabilizer state $\ket{\Psi_{s,1}}^{(N)} = \ket{1}^{\otimes N}$.

In the regime well above the phase transition $\vbar_x \gg 1$, the system is in the deformed phase (the mean field solution spontaneously breaks the parity symmetry) with a maximal deformation reached for $\vbar_x \gg N $. In this region the exact solution tends to exhibit maximal entanglement (von Neumann entropy and $8$-tangle) and low magic, and thus can be captured to a large extent by the entangled stabilizer state $\ket{\Psi_{s,2}}^{(N)}$.

We observe that, independently of the size $N$ of the system, the exact ground state displays (close) to maximal $N$-tangle $\tau_N$, while the $n$-tangles $\tau_n$ with $n<N$ are several orders of magnitude smaller, and negligible in comparison. The system thus mostly exhibits genuine collective $N$-partite entanglement.
Interestingly, while the entangled stabilizer state $\ket{\Psi_{s,2}}^{(N)}$ displays constant maximal $\tau_N$, we find that $\tau_n = 0$ for $n<N$. This confirms that such stabilizer state is able to capture, to a large extent, both bipartite and multipartite entanglement properties of the exact solution in the region $\vbar_x \gg 1$.

It is the region in between, around $\vbar_x \simeq 2$, where both single-particle and collective effects come into play, that is most difficult to describe with a single stabilizer state. As $\vbar_x \simeq 2$ is the point where the stabilizer ground state transitions from an unentangled to an entangled one, this is also the region where the exact ground state is furthest described by either stabilizer state, and for that reason, the magic is maximal around this point.

\begin{figure}[t]
     \centering
        \includegraphics[width=\columnwidth]{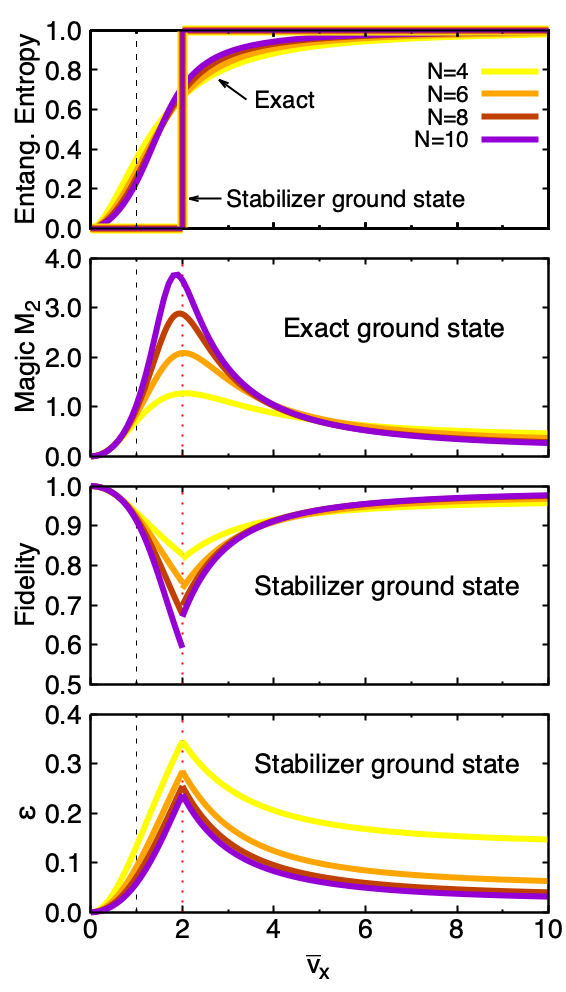}
         \caption{From bottom to top panel: relative energy difference between exact and stabilizer ground-state energy, fidelity of the stabilizer ground state, stabilizer 2-R\'enyi entropy $\mathcal{M}_2$ of the exact ground state, entanglement entropy of both exact and stabilizer ground states. The results are shown for various system sizes and have been obtained with $\chi= -1$.
         The stabilizer ground state is taken to be $\ket{\Psi_{s,1}}^{(N)}$ for $\vbar_x < 2$ and $\ket{\Psi_{s,2}}^{(N)}$ for $\vbar_x > 2$. }
         \label{fig:Stabil_gs_Ns}
\end{figure}

We note that this transition point at $\vbar_x \simeq 2$ is based on the energy criterion chosen to select the stabilizer ground state (Sec.~\ref{sec:general} above) and coincides with the fidelity crossing point for the present chosen example ($N=8$ and $\chi=-1$). Depending on the parameter values, however, the fidelity crossing point can occur at slightly lower values of $\vbar_x$ (see Figs.~\ref{fig:Stabil_gs_Ns}, \ref{fig:mag_Jzop_chi-0p1}, and appendix~\ref{sec:app_stabil_Hami}). \\

Fig.~\ref{fig:Stabil_gs_Ns} shows fidelity and relative energy difference for the stabilizer ground state, which corresponds to $\ket{\Psi_{s,1}}^{(N)}$ for $\vbar_x < 2$ and $\ket{\Psi_{s,2}}^{(N)}$ for $\vbar_x > 2$, according to the energy minimization criterion of Sec.~\ref{sec:general}, for system sizes from $N=4$ to $N=10$.
The entanglement entropy $S^{(N)}_1$ and SRE $\mathcal{M}_2(\ket{\Psi})$ of the exact solution are also provided.
It is clear that the magic is extensive in the region around the stabilizer ground state transition point $\vbar_x \simeq 2$. This was also noted in Ref.~\cite{Passarelli:2024tyi}, for the case $\chi =0$. This is accordance with the behavior of the fidelity of the stabilizer ground state shown in the middle panel, which decreases in this region for large $N$. 
On the other hand, the relative energy difference behaves similarly to the entanglement entropy, {\it i.e.} it tends to improve and converges as $N$ increases.
%

%%%%%%%%%%%%%%%%%%%%%%%%%%%%%%%%%%%%%%%%%%
%%%%%%%%%%%%%%%%%%%%%%%%%%%%%%%%%%%%%%%%%%
%%%%%%%%%%%%%%%%%%%%%%%%%%%%%%%%%%%%%%%%%%
\section{Beyond Stabilizer Ground State with Magic Injection}
\label{sec:Beyond_stabilizer}
%%%%%%%%%%%%%%%%%%%%%%%%%%%%%%%%%%%%%%%%%%
%%%%%%%%%%%%%%%%%%%%%%%%%%%%%%%%%%%%%%%%%%
%%%%%%%%%%%%%%%%%%%%%%%%%%%%%%%%%%%%%%%%%%

%%%%%%%%%%%%%%%%%%%%%%%%%%%%%%%%%%%%%%%%%%
%%%%%%%%%%%%%%%%%%%%%%%%%%%%%%%%%%%%%%%%%%
\subsection{Discussion and Preliminaries}
%%%%%%%%%%%%%%%%%%%%%%%%%%%%%%%%%%%%%%%%%%
%%%%%%%%%%%%%%%%%%%%%%%%%%%%%%%%%%%%%%%%%%

Once the stabilizer ground state has been obtained, there are several ways that one can adopt to incorporate non-stabilizerness into the system.

For example, one can in principle apply a unitary operator $\hat{U}(\theta) = e^{-i \theta \hat O}$, where $\theta$ is a continuous angle that can be determined variationally.
There is a range of one- and two-spin Hermitian operators $\hat O$ that one can consider to generate such unitary.
For example, one can use terms of the Hamiltonian itself, as done in Hamiltonian variational ansatz (see {\it e.g.}~\cite{PhysRevA.92.042303}), and/or commutators of Hamiltonian terms, to explore a larger part of the Hilbert space~\cite{Farrell:2023fgd}, as well as individual fermionic or qubit excitation operators~\cite{Grimsley:2018wnd,Tang:2019tpm}.

In the present case of the LMG model, employing $\hat O = \hat{J}_y$ on top of the unentangled state $\ket{\Psi_{s_1}} = \ket{1}^{\otimes N}$, corresponds to generating the deformed (unresticted) HF solution~\footnote{In fact, the deformed HF solution can also be interpreted as the stabilizer ground state of the transformed Hamiltonian $\widetilde{H} (\theta) = \hat{U}^\dagger(\theta) \hat H \hat{U}(\theta)$. We have indeed checked that the transformed Hamiltonian does not admit an entangled stabilizer ground state.}. 
As mentioned previously, the downside of this procedure is that it breaks parity symmetry, which then has to be restored {\it a posteriori}. While such restoration is straightforward in the case of the LMG model, in realistic fermionic systems, such as atomic nuclei, this procedure is numerically highly costly.

In fact, the only Hermitian excitation operators which would preserve both the reality and symmetries (parity and permutation) of the wave function, are 
$ \hat{O}_\pm = J_x J_y \pm J_y J_x 
= \frac{1}{2}\sum_{ij} (X_i Y_j \pm Y_i X_j) 
$.
The individual terms in $\hat{O}_\pm$, however, do not commute, and thus implementing this operator on a quantum computer typically requires Trotterization for large values of $N$, which results in approaches similar to as Trotterized UCC~\cite{D1CS00932J} or ADAPT-VQE~\cite{Grimsley:2018wnd} ansatz. 
In appendix~\ref{sec:app_adapt} we have implemented ADAPT-VQE using the stabilizer ground state as initial state.
In the regime of large deformation/collectivity, we find that such strategy does not appear to be optimal, as the individual excitations somewhat destroy the previously-build coherence of the entangled stabilizer state, before rebuilding it layer by layer. As illustrated in appendix~\ref{sec:app_adapt}, this induces long plateaux in the convergence procedure, which appear to be characteristic of collective initial states, due to equally small energy gradients of the individual operators for a large number of iterations. Only after many layers, once the collectivity has been re-built, these gradients increase again and convergence resumes.
Such plateaux caused by gradients troughs~\cite{Grimsley:2022azc} have been noticed in the case of symmetry-breaking reference states, for example in e.g. Refs~\cite{Bertels:2022,Tsuchimochi:2022ucu}, in or Ref.~\cite{Romero:2022blx} when using a LMG deformed HF solution as initial state in ADAPT-VQE. In the present work, we also observe such feature although the initial stabilizer state is symmetry-preserving.
\\

In the following we explore a different strategy, to preserve the collectivity of the system together with the symmetries along the variational process, making use of imaginary-time evolution techniques.
Specifically, we apply both a "variational" and standard imaginary time evolution, as described below.

%%%%%%%%%%%%%%%%%%%%%%%%%%%%%%%%%%%%%%%%%%
%%%%%%%%%%%%%%%%%%%%%%%%%%%%%%%%%%%%%%%%%%
\subsection{Variational Quantum Imaginary Time Propagation}
\label{sec:Var_QITP}
%%%%%%%%%%%%%%%%%%%%%%%%%%%%%%%%%%%%%%%%%%
%%%%%%%%%%%%%%%%%%%%%%%%%%%%%%%%%%%%%%%%%%

The first approach is based on the observation that
the entangled stabilizer state $\ket{\Psi_{s,2}}^{(N)}$ contains all the configurations that contribute to the exact ground state (and only those) with equal amplitude probabilities. In order to arrive at the exact solution, the task is therefore to adjust the amplitudes related to these configurations.
This can be done, to a large extent, by application of the non-unitary operation 
\begin{align}
    \ket{\Phi (\theta)} 
    & = \text{e}^{- \theta \hat{J}_z} \ket{\Psi_{s,2}}^{(N)} \; , \nonumber \\
    & = \prod_{i=1}^N \text{e}^{- \theta Z_i} \ket{\Psi_{s,2}}^{(N)} \; ,
\label{eq:var_QITP_Jz}
\end{align}
where the angle $\theta$ can be variationally optimized. 
\\

\begin{figure}[h]
     \centering
         \includegraphics[width=\columnwidth]{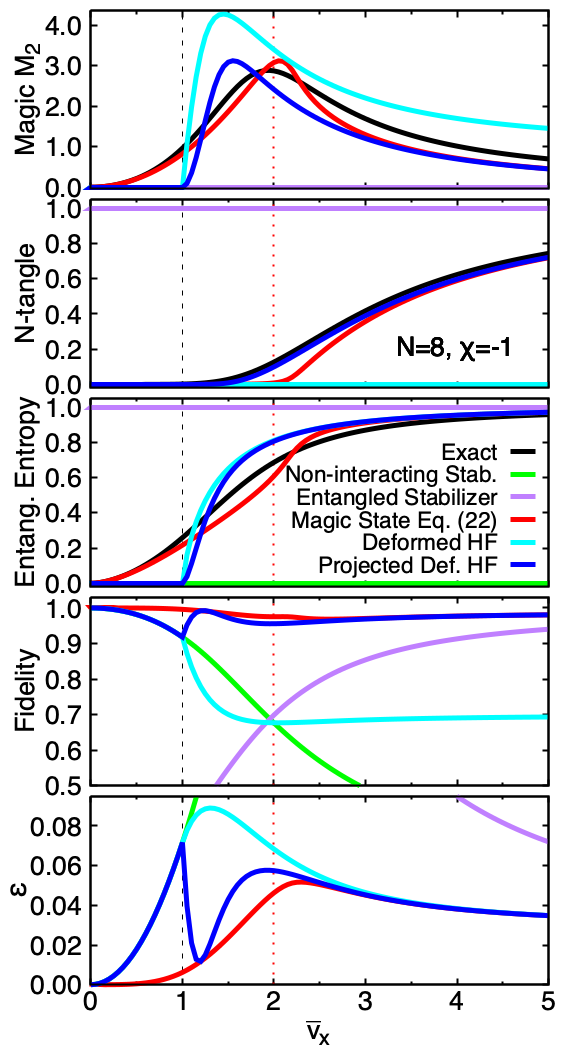}
         \caption{Relative energy difference $\varepsilon$, fidelity, von Neumann entanglement entropy $S_1^{(N)}$, $N$-tangle $\tau_N$, and SRE $\mathcal{M}_2$ in a system with $N=8$ spins, and $\chi=-1$. The exact solution appears in black. The unentangled and entangled stabilizer states, $\ket{\Psi_{s,1}}^{(N)}$ and $\ket{\Psi_{s,2}}^{(N)}$, are shown in green and purple curves, respectively. The state obtained via Eq.~\eqref{eq:var_QITP_Jz} is shown with red curves. For comparison, the deformed HF without and with projection are shown with cyan and blue curves, respectively. }
         \label{fig:mag_Jzop_chi-1}
\end{figure}

\begin{figure}[h]
     \centering
         \includegraphics[width=\columnwidth]{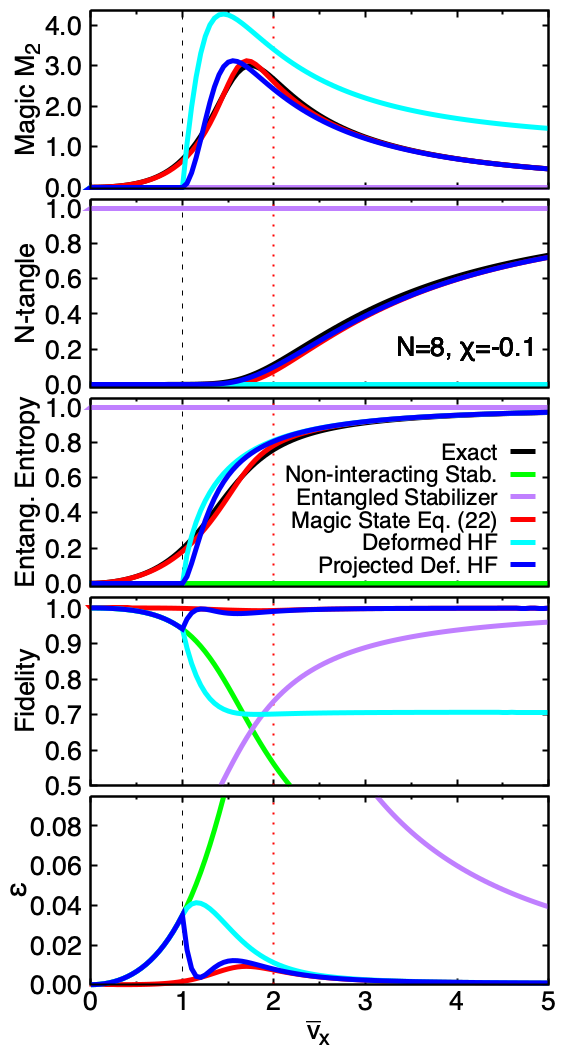}
         \caption{Same as Fig.~\ref{fig:mag_Jzop_chi-1} for $\chi=-0.1$. }
         \label{fig:mag_Jzop_chi-0p1}
\end{figure}

We show in Fig.~\ref{fig:mag_Jzop_chi-1} and \ref{fig:mag_Jzop_chi-0p1} the results obtained for a system of $N=8$ spins, using parameter values $\chi=-1$ and $\chi=-0.1$, respectively. 
In practice, the optimal value of $\theta$ has been determined by performing an initial scan of the energies obtained for a grid of $\theta$ values. The retained value was then further refined using optimizers available in Scipy~\cite{2020SciPy-NMeth}. Specifically, we used both COBYLA and gradient based L-BFGS-B optimizers which provided results in agreement with each other.

The magic state obtained via Eq.~\eqref{eq:var_QITP_Jz} is shown with a red curve, and recovers the exact solution to a large extent, even in the region around and below the phase transition at $\vbar_x=1$. We observe that the results are more accurate for smaller absolute values of $\chi$, since the $YY$ term in the Hamiltonian has smaller weight, and thus, applying operator $\text{exp}(- \theta \hat{J}_z)$ to the stabilizer state $\ket{\Psi_{s,2}}^{(N)}$ captures the exact solution almost fully. In all cases, the largest deviations to the exact result appear around $\vbar_x =2$, where the magic presents a maximum.
\\

For information, we also provide the deformed Hartree-Fock solution without and with projected onto good parity, with cyan and blue colors, respectively.
As mentioned above, the deformed HF solution is obtained via global single-spin transformation of the non-interacting ground state $\ket{1}^{\otimes N}$ around the $y$ axis. As it is a simple mean field state, the deformed HF solution features the same computational complexity as a stabilizer state and can be prepared efficiently classically. 
This state has vanishing magic and entanglement in the rotated spin basis, however SRE and entanglement entropy become non-zero if calculated in the original (non-rotated) basis. This is what is shown in Figs.~\ref{fig:mag_Jzop_chi-1} and \ref{fig:mag_Jzop_chi-0p1}. 
We note however that the HF state is not able to capture the $n$-tangles, which are independent of the qubit (spin) basis~\cite{Hengstenberg:2023ryt} and thus remain equal to zero before projection~\footnote{This is because the basis transformation is a qubit rotation around the $y$ axis and thus commutes with the $Y$ Pauli operators in the $n$-tangles. The expectation value in Eq.~\eqref{eq:tangles} therefore remains invariant under such transformation}.
Additionally, although the deformed HF state describes the energy of the system to a better extent than the stabilizer ground state, the fidelity of the wave function is not well reproduced in the region of large deformation (large $\vbar_x$), and requires projection.

While the projection largely improves the results above the phase transition, the resulting solution displays a discontinuous behavior at the phase transition $\vbar_x=1$.
This is not the case with the strategy of Eq.~\eqref{eq:var_QITP_Jz} which displays a smooth behavior throughout the full range of $\vbar_x$ values, and provide results that are overall at least as good as the projected HF solution, and of better quality around the phase transition.
As we have verified,  the remaining discrepancies around $\vbar_x =2$ can in principle be systematically improved using higher order operators, for example, via $\ket{\Phi (\theta_1, \theta_2)} = \text{exp}(- \theta_2 \hat{J}_z^2)  \, \text{exp}(- \theta_1 \hat{J}_z) \, \ket{\Psi_{s,2}}^{(N)}$. 
\\

In a quantum circuit, the non-unitary operator in Eq.~\eqref{eq:var_QITP_Jz} can be implemented, for example, using existing techniques developed for Quantum Imaginary Time Evolution (QITE) such as the Quantum Imaginary Time Propagation (QITP) algorithm developed in Ref.~\cite{Turro:2021vbk}. 
These techniques are inspired by classical imaginary-time evolution (ITE) methods, which, in their original form, are used to find the ground state of a physical system, by evolving an initial state (with non-zero overlap with the ground state) in imaginary time, so that after a long time, the system converges to its exact ground state.
By making use of an ancillary system, QITP allows the implementation of the non-unitary ITE operator on a quantum circuit, by acting with a carefully chosen unitary on the full extended system (see details in Ref.~\cite{Turro:2021vbk} which are summarized in appendix~\ref{sec:app_QITP}).
This method can be adapted to implement Eq.~\eqref{eq:var_QITP_Jz}, which represents some kind of ITE with only part of the Hamiltonian ($\hat{J}_z$ operator), and where the angle $\theta$ plays the role of imaginary time. Contrarily to ITE, there is an optimal value $\theta=\theta_{opt}$ to be found, which provides a minimal energy. Such optimal value can be determined via standard optimization procedures as above.  A comparison between the exact evolution of Eq.~\eqref{eq:var_QITP_Jz} (shown in Figs.~\ref{fig:mag_Jzop_chi-1} and \ref{fig:mag_Jzop_chi-0p1}) and the evolution implemented via variational QITP is provided in appendix~\ref{sec:app_QITP}.

%%%%%%%%%%%%%%%%%%%%%%%%%%%%%%%%%%%%%%%%%%
%%%%%%%%%%%%%%%%%%%%%%%%%%%%%%%%%%%%%%%%%%
\subsection{Full Quantum Imaginary Time Propagation}
\label{sec:Full_QITP}
%%%%%%%%%%%%%%%%%%%%%%%%%%%%%%%%%%%%%%%%%%
%%%%%%%%%%%%%%%%%%%%%%%%%%%%%%%%%%%%%%%%%%
While the magic-injection procedure described above allows for recovering the exact solution to a large extent in the case of the LMG model, it is not clear whether it can be successfully applied to a more realistic system~\footnote{Typically we expect that this procedure would only be useful if the degree of collectivity of the system is large (many components with similar weights), and the corresponding stabilizer ground state contains most relevant components to the exact ground state.}. 
In the case of a general Hamiltonian, implementations of QITE/QITP with the full Hamiltonian may be employed to inject magic. In general, the advantage of this procedure is that it does not require a particular ansatz for the evolved quantum state, avoids the need for an optimization procedure, and, in principle, guarantees convergence to the exact ground state. On the other hand, the drawback is that the size of the qauntum circuit may increase rapidly.
In this context, since the convergence towards the ground state, and thus the size of the quantum circuit, are governed by the overlap of the initial state with the exact ground state, it is highly desirable to have an initial state that captures features of the exact ground state to a large extent. This motivates exploring the use of stabilizer states as starting points for QITE/QITP. 
Thus, in order to gain insight for future studies of more general Hamiltonians, we investigate below the use of stabilizer states as initial states for QITP with the full LMG Hamiltonian.
Details of the procedure are reminded in appendix~\ref{sec:app_QITP}.

We show in Fig.~\ref{fig:full_QITP} the fidelity $|\langle \Phi_{ex} | \eta(\tau) \rangle|$ between the exact ground state and the state $\ket{\eta (\tau)}$ evolved with the QITP operator (see Eq.~\eqref{eq:QITE_final_state} in appendix~\ref{sec:app_QITP}) as a function of the imaginary time $\tau$. We compare the cases when the initial state is chosen to be the non-interacting (unentangled) state $\ket{\Psi_{s,1}}^{(N)}$ and the entangled stabilizer state $\ket{\Psi_{s,2}}^{(N)}$.
We choose different values of the interaction strength near and away from the phase transition, and the point of stabilizer ground state transition, specifically, $\vbar_x = 1.1, \; 2.1, \; 5.0, \; 10.0$.
\begin{figure}
     \centering
         \includegraphics[width=\columnwidth]{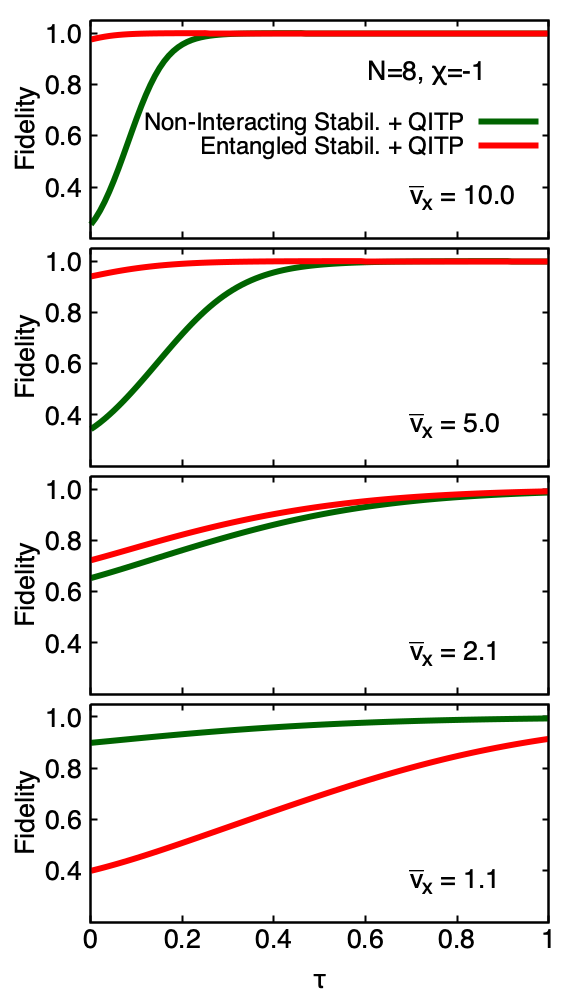}
         \caption{Fidelity $|\langle \Phi_{ex} | \eta(\tau) \rangle|$ of the state $\eta(\tau)$ evolved via QITP operator (defined in Eq.~\eqref{eq:QITE_final_state}) for two initial states $\ket{\eta(0)} = \ket{\Psi_{s,1}}^{(N)}$ (green curve) and $\ket{\eta(0)} = \ket{\Psi_{s,2}}^{(N)}$ (red curve), as a function of $\tau$. The results are shown for a system of $N=8$ spins and $\chi=-1$, for different values of the interaction strength $\vbar_x$. }
         \label{fig:full_QITP}
\end{figure}
As expected, in the region $\vbar_x < 2$ the unentangled state provides a better initial state due to higher fidelity with the exact solution, while the entangled one becomes optimal for $\vbar_x > 2$ as $\vbar_x$ grows. The convergence towards the exact solution also appears to be more rapid away from the phase transition, for large values of $\vbar_x$. Overall the results indicate that using stabilizer ground states for subsequent quantum or classical ITE methods may be optimal, in particular in systems exhibiting a large degree of collectivity, due to the large-scale entanglement captured by the stabilizer ground state. 
As mentioned above, this is particularly important in the quantum case as the success of QITE/QITP algorithms directly depends on the overlap of the initial state with the exact ground state, and often requires amplitude amplification~\cite{595153,Brassard_2002}, leading to an additional increase of the size of the quantum circuit. The use of an entangled stabilizer state thus may alleviate this need in the case of collective systems, and is to be investigated in more details in the future.

Note that, in contrast with the variational QITP described in Sec.~\ref{sec:Var_QITP}, full QITP will in practice require Trotterization. Studying the effect of Trotterization together with stabilizer ground states is left for a future study.

\section{Summary and Conclusions}
\label{sec:conclu}

In this work we have made use of the stabilizer formalism in order to approximate the ground state of a quantum many-body system with collective behavior. Specifically we have chosen the Lipkin-Meshkov-Glick (LMG) model which can be described by a system of $N$ interacting spins with all-to-all connectivity, 
and displays a second-order phase transition 
to a symmetry-broken state, corresponding to a collective regime where all spins are correlated in a coherent manner, and analogous of the transition between spherical and deformed phases in atomic nuclei.

The procedure for finding the stabilizer ground state adopted in this work is based on a decomposition of the Hamiltonian into mutually-commuting operators forming a stabilizer part, and a residual part inducing non-stabilizerness (magic). The optimal stabilizer Hamiltonian is determined via an energy minimization criterion, and the corresponding stabilizer ground state is efficiently prepared in terms of a graph state, using an existing technique based on stabilizer tableaux.
We found that the resulting stabilizer ground state coincides with the non-interacting (unentangled) ground state in the normal phase while it becomes maximally entangled in the region of large deformation (collectivity), both in the bi-partite and $N$-partite sense, which captures to a large extent 
the entanglement features of the exact solution.
The transition point between unentangled and entangled stabilizer ground states coincides with the maximum of magic, as it is the region that is most difficult to capture with a single stabilizer state.

This study now motivates the use of the techniques proposed in this work to describe more complex and realistic systems.
Ultimately stabilizer states could potentially provide a path towards efficient simulations of systems with high degrees of collectivity from an {\it ab initio} single-particle picture, without resorting to traditional explicit symmetry breaking and restoration techniques. 

To compare these different approaches from the perspective of quantum complexity relative to a particular quantum resource theory, the stabilizer-plus-magic formalism of the present paper relates to the resource theory based on Clifford-plus-T-gates, while the more traditional many-body techniques based on (symmetry-broken) mean-field-plus-fluctuations relate to the resource theory where fermionic Gaussian unitaries ("matchgates") are the free resources and non-Gaussianity is the costly one, see {\it e.g.} Ref.~\cite{Sierant:2025fax}. Although this requires further investigations, this work seems to indicate that shape collectivity could be efficiently reached within the former resource theory, while the latter one would lead to longer paths in the Hilbert space, ultimately reached via projection, and thus necessitates more resources.

While the method described in this work provides a systematic way of determining stabilizer ground states in various many-body systems, the best path for injecting magic on top of such stabilizer ground state, while preserving collectivity and symmetries, is not as clear, and will require specific investigations depending on the nature of the system of interest and the computing device (classical, quantum or hybrid).
Specifically, if the residual magic interaction $W$ is “small” one can incorporate it via perturbation theory or fluctuations on top of the stabilizer ground state, see Ref.~\cite{Gu:2024ure}. If it is non-perturbative, one can apply powerful classical methods, such as, for instance, coupled cluster, on top of the stabilizer state, or, at scale, quantum algorithms implemented on quantum computers.
For example, in the present study of the LMG model, we found that ADAPT-VQE with few-body excitation operators 
does not appear to be adequate when applied on collective stabilizer ground state as it tends to destroy 
the coherence of the state, ultimately requiring a similar number of iterations to converge than with a trivial unentangled state. 
On the other hand, the use of stabilizer ground state promises to substantially accelerate classical or quantum imaginary-time evolution methods, due to larger overlaps with the exact ground states.

In systems such as atomic nuclei, which often present a strong interplay between collective and single-particle degrees of freedom, we found in a previous study that (the onset of) shape collectivity is also associated with large magic~\cite{Brokemeier:2024lhq}. This is consistent with the present work which suggests that collectivity is associated with large-scale entanglement while the interplay with the single-particle regime produces magic.
The magic injection is therefore expected to be crucial in these systems. In this context, it will be interesting to investigate whether stabilizer states may be more efficient in spreading magic than typically-used unentangled reference states, and the use of $t$-doped quantum circuits~\cite{OLIVIERO2021127721,Leone_2021}, alternating between Clifford and magic gates, may be useful.
This will be studied in a subsequent work.
In the future, we also plan to investigate the combination of symmetry-preserving basis transformations (or Hamiltonian transformations) to decrease the computational complexity, together with
the present stabilizer ground state method via extensions of our previously-developed HL-VQE algorithm.

%%%%%%%%%%%%%%%%%%%%%
\begin{acknowledgements}
%%%%%%%%%%%%%%%%%%%%%
I would like to thank Martin J. Savage for providing feedback on the manuscript, as well as for many inspiring discussions and previous collaborations which led me to this work. 
I also thank Denis Lacroix for useful discussions, in particular about ADAPT-VQE.
I would like to further acknowledge several exchanges during the workshop {\it Entanglement in Many-Body Systems: From Nuclei to Quantum Computers and Back}~\footnote{\url{https://iqus.uw.edu/events/entanglementinmanybody/}} held at the InQubator for Quantum Simulation (IQuS), and during the {\it First Workshop on Many-Body Quantum Magic} (MBQM24)~\footnote{\url{https://mbqm.tii.ae/talks.php}} held at the Technology Innovation Institute.
This work was supported by Universit\"at Bielefeld and by ERC-885281-KILONOVA Advanced Grant.
\end{acknowledgements}

\vspace{0.1cm}
Data availability: 
The numerical results plotted in the figures of this article are available at Ref.~\cite{3003952}.

\bibliography{biblio_paper}

%%%%%%%%%%%%%%%%%%%%%
%%%%%%%%%%%%%%%%%%%%%
\onecolumngrid
\appendix
%%%%%%%%%%%%%%%%%%%%%
%%%%%%%%%%%%%%%%%%%%%

\section{Stabilizer Hamiltonian, Stabilizer Ground States, and Relation to Graph States}
\label{sec:app_stabil_Hami}

In this section we provide details on the procedure to find the stabilizer Hamiltonian and prepare the corresponding stabilizer ground state of the LMG model. 
As a reminder, the full Hamiltonian of this system is:
\begin{align}
    \widetilde{H} 
    & = \hat{J}_z - \frac{\bar{v}_x}{(N-1)} (\hat J_x^2 + \chi \hat J_y^2) \; ,\nonumber \\
    & = \frac{1}{2} \sum_i Z_i -  \frac{\bar{v}_x}{2(N-1)} \sum_{i<j}( X_i X_j + \chi Y_i Y_j) \; ,
    \label{eq:Hami_app}
\end{align}
As in the main text, we use the convention
\begin{align}
\ket{\uparrow} = \ket{0} = 
    \begin{pmatrix}
        1 \\ 0
    \end{pmatrix}
    \text{    and    }
    \ket{\downarrow} = \ket{1} = 
    \begin{pmatrix}
        0 \\ 1
    \end{pmatrix}
    \; ,
\end{align}
so that the non-interacting ground state of the LMG model (obtained for $\vbar_x=0$) is the state 
$\ket{\downarrow \downarrow \downarrow ... \downarrow} = \ket{111..1} $.
\\
We will first consider the case $\chi = -1$, and generalize to $\chi \in [-1,0)$ in a second stage.

\subsection{Two-spin system (N=2)}
The LMG Hamiltonian for the two-spin system with $\chi = -1$ is 
\begin{align}
    \widetilde{H}^{(2)} 
    = \frac{1}{2} (Z_1 + Z_2) -  \frac{\vbar_x}{2} ( X_1 X_2- Y_1 Y_2) \; ,
\end{align}
so that the set of Pauli string in $\widetilde{H}^{(2)}$ is
\begin{align}
    \mathcal{G}_H = \{ Z_1, Z_2, X_1 X_2, Y_1 Y_2\}  \; ,
\end{align}
from which we extract two sets of commuting operators:
\begin{align}
Q_1 = \{ Z_1, Z_2\} \text{ and } Q_2 = \{ X_1 X_2 , Y_1 Y_2\} \; .
\end{align}
Each of these sets already contains $N=2$ generators, generating stabilizer groups with $2^N = 4$ elements. 
We need to examine all sign possibilities, i.e. consider 
\begin{align}
    Q_1^a &= \{ Z_1, Z_2\}  \; , \\
    Q_1^b &= \{ -Z_1, Z_2\}  \; , \\
    Q_1^c &= \{ Z_1, -Z_2\}  \; , \\
    Q_1^d &= \{ -Z_1, -Z_2\}  \; , 
\end{align}
as well as 
\begin{align}
    Q_2^a &= \{ X_1 X_2 , Y_1 Y_2\}  \; , \\
    Q_2^b &= \{ -X_1 X_2 , Y_1 Y_2\}  \; , \\
    Q_2^c &= \{ X_1 X_2 , -Y_1 Y_2\}  \; , \\
    Q_2^d &= \{ -X_1 X_2 , -Y_1 Y_2\}  \; .
\end{align}
Each of these $Q_i^{k}$ generates a stabilizer group $\mathcal{S}_i^k = \langle Q_i^{k} \rangle$ each of which corresponds to a unique stabilizer state.
\\

The corresponding stabilizer energy for each of the  $Q_1^{k}$ groups are
\begin{align}
    E(Q_1^a) &= \frac{1}{2}  + \frac{1}{2}  = 1  \; ,\\
    E(Q_1^b) &= -\frac{1}{2}  + \frac{1}{2}  = 0  \; ,\\
    E(Q_1^c) &= \frac{1}{2}  - \frac{1}{2}  = 0  \; ,\\
    E(Q_1^d) &= -\frac{1}{2}  - \frac{1}{2}  = -1  \; .
\end{align}
It is easy to guess that the corresponding stabilizer states are $\ket{00}$, $\ket{10}$, $\ket{01}$, $\ket{11}$, respectively, which are obviously non-entangled. Among these, $\ket{11}$ is the one with the lowest energy, corresponding to the non-interacting ground state.
We also have the following stabilizer energies for the $Q_2^k$ groups:
\begin{align}
    E(Q_2^a) &= - \frac{\bar{v}}{2} (1 -1) =  0  \; ,\\
    E(Q_2^b) &= - \frac{\bar{v}}{2} (-1 -1)  = \vbar_x  \; ,\\
    E(Q_2^c) &= - \frac{\bar{v}}{2} (1 +1) = -\vbar_x  \; ,\\
    E(Q_2^d) &= - \frac{\bar{v}}{2} (-1 +1)  = 0 \; .
\end{align}
Similarly, it is clear that the stabilizer states corresponding to the $\mathcal{S}_2^k = \langle Q_2^{k} \rangle$ stabilizer groups are Bell states which are known to be simultaneous eigenvectors of $XX, YY$ (and $ZZ$). 
Among them, the stabilizer state with the lowest energy is the symmetric spin-aligned Bell state $(\ket{00} + \ket{11})/\sqrt{2}$ with energy $- \vbar_x$ (since $\vbar_x \geq 0$).
Note that the stabilizer state stabilized by $\mathcal{S}_2^b$, with energy $+\vbar_x$ is the antisymmetric Bell state $(\ket{00} - \ket{11})/\sqrt{2}$. The others are the spin anti-aligned states $(\ket{01} \pm \ket{10})/\sqrt{2}$.
\\

In summary, in the case of a two-spin system: for $\vbar_x < 1$, the stabilizer ground state is the non-interacting state $\ket{\Psi_{s,1}} \equiv \ket{11}$ with energy $E_{s,1} = -1$, while for $\bar{v} > 1$ the stabilizer ground state is the entangled Bell state $\ket{\Psi_{s,2}} = (\ket{00} + \ket{11})/\sqrt{2}$ with energy $E_{s,2} = - \bar{v}_x$.
\\

The unentangled state $\ket{\Psi_{s,1}}$ is trivially prepared from $\ket{00}$ with X gates, while the Bell state $\ket{\Psi_{s,2}}$ can be prepared with a single Hadamard and CX gate:
\begin{align}
    \ket{\Psi_s}^{(2)} 
    = \frac{1}{\sqrt{2}} \left(\ket{00} + \ket{11} \right)
    = \text{CX}_{21}  \ket{0} \otimes \ket{+}
    = \text{CX}_{21} \text{H}_2 \ket{0} \otimes \ket{0}   \; ,
\label{eq:state_prep_N2}
\end{align}
where $\text{CX}_{ji}$ is controlled by qubit $j$ and has qubit $i$ as target.

\subsection{Three-spin system (N=3)}
The Hamiltonian for $N=3$ is 
\begin{align}
    \widetilde{H}^{(3)} 
    = \frac{1}{2} (Z_1 + Z_2 + Z_3) -  \frac{\bar{v}}{4} ( X_1 X_2 + X_1 X_3 + X_2 X_3 - Y_1 Y_2 - Y_1 Y_3 - Y_2 Y_3) \; .
\end{align}
Similarly as above we can consider the subsets of operators consisting only of $Z$ gates:
\begin{align}
    Q_1^k = \{ \pm Z_1, \pm Z_2, \pm Z_3\} \; ,
\end{align}
where the superscript $k$ denotes the possible sign combinations.
These subsets generate stabilizer groups $\mathcal{S}_1^k = \langle Q_1^k \rangle$, each with 8 elements.
We can also extract groups with XX and YY operators. Contrarily to the $N=2$ case, there are now different ways of grouping these operators into commuting subsets. For example one can consider:
\begin{align}
    Q_2^k &= \{ \pm X_1 X_2 , \pm X_1 X_3, \pm X_2 X_3 \}  \; , \\
    Q_3^k &= \{ \pm Y_1 Y_2 , \pm Y_1 Y_3, \pm Y_2 Y_3 \} \; , \\
\end{align}
or, alternatively,
\begin{align}
    Q_4^k &= \{ \pm X_1 X_2 , \pm Y_1 Y_2, \pm Z_3\}  \; ,\\
    Q_5^k &= \{ \pm X_1 X_3 , \pm Y_1 Y_3, \pm Z_2\}  \; ,\\
    Q_6^k &= \{ \pm X_2 X_3 , \pm Y_2 Y_3, \pm Z_1\} \; . \\
\end{align}

\begin{itemize}

\item Among the $Q_1^k$ operators the minimum stabilizer energy is obtained for $\{ -Z_1,  -Z_2,  -Z_3\} $ which again corresponds to the state $\ket{111} $, with energy $-3/2$.

\item Among the $Q_4^k$ operators the minimum stabilizer energy is given by $-1/2 -\bar{v}/4 (1+1) = -1/2 -\bar{v}/2$, and is the same for $Q_5^k$, $Q_6^k$. Note that picking one of these groups would lead to a stabilizer state breaking permutation symmetry.

\item Among the $Q_2^k$ operators the minimum stabilizer energy is equal to $(-\bar{v}/4) (1+1+1) = -3\vbar_x/4 $, which corresponds to $\{ + X_1 X_2 ,  + X_1 X_3,  + X_2 X_3\}$.
Out of these three operators, two are linearly independent, for example, $+ X_1 X_3$ and  $+ X_2 X_3$. Thus one needs to complete this set with a third operator to generate a complete stabilizer group and specify the corresponding stabilizer state uniquely. 
We require that the stabilizer state satisfies both permutation and parity symmetry. In particular, it should have an odd number of spins down, since, according to the form of the Hamiltonian, the non-interacting ground state $\ket{111}$ can only couple to states with odd number of spins down. An operator which satisfies these conditions is $- Z_1 Z_2 Z_3$.
It is easy to see that the state
\begin{align}
    \ket{\Psi_{s,2}}^{(3)} =  \frac{1}{2} (\ket{111} + \ket{100} + \ket{010} + \ket{001})  \; ,
\end{align}
which is a superposition of the 3-qubit W state and $\ket{111}$, is stabilized by 
$\mathcal{S} = \langle + X_1 X_3, + X_2 X_3, - Z_1 Z_2 Z_3 \rangle$ and satisfies the required symmetries.

\item For $Q_3^k$,  the minimum stabilizer energy is equal to $-\vbar_x/4 $, which corresponds to, for example, $\{+Y_1 Y_2 ,  -Y_1 Y_3,  -Y_2 Y_3\}$ and, e.g., $\mathcal{S} = \langle  -Y_1 Y_3,  -Y_2 Y_3, -Z_1Z_2Z_3\}$~\footnote{Note that it is not possible to have a group generating $-Y_iY_j$, with minus signs for all $i<j$ for $N\geq 3 $.}. The corresponding stabilizer ground state is similar to $\ket{\Psi_{s,2}}^{(3)}$ with different relative signs.

\end{itemize}

In summary, for a system with $N=3$ spins: if $ -3\vbar_x/4 > -3/2 $, i.e. $\vbar_x < 2$, the stabilizer ground state is the unentangled stabilizer state $\ket{\Psi_{s,1}}^{(3)} = \ket{111}$ with energy $E_{s,1}=-3/2$, while for $\vbar_x>2$ the stabilizer ground state is the entangled state $\ket{\Psi_{s,2}}^{(3)}$ with energy $E_{s,2}= -3\bar{v}_x/4 < -3/2$. The permutation-symmetry breaking choice never leads to a stabilizer ground state.
\\
It is straightforward to see that the entangled state $\ket{\Psi_{s,2}}^{(3)}$ can be prepared with a series of CNOT, H and X gates:
\begin{align}
    \ket{\Psi_s}^{(3)}   
    &= \frac{1}{2} \left(\ket{100} + \ket{111} + \ket{010} + \ket{001} \right)  \; , \nonumber \\
    & = \text{CX}_{21} \text{CX}_{31} \text{CX}_{32} \ket{1} \otimes \ket{+} \otimes \ket{+}   \; ,\nonumber \\
    & = \text{CX}_{21} \text{CX}_{31} \text{CX}_{32} \text{X}_1 \text{H}_2 \text{H}_3 \ket{0} \otimes \ket{0} \otimes \ket{0} 
    \; ,
\label{eq:state_prep_N3}
\end{align}

\subsection{Arbitrary $N>2$}

The procedure for extracting subsets of mutually-commuting operators in the $N=3$ case can be generalized to arbitrary values of $N$.

The stabilizer group $\mathcal{S}_1^{(N)}$ generated by $\{ -Z_1, -Z_2, ... -Z_N \}$ operators has stabilizer energy $E_{s,1}= -N/2$ and always coincides with the non-interacting ground state $\ket{\Psi_{s,1}}^{(N)} = \ket{1}^{\otimes N}$.

As for $N=3$, the interacting part of the Hamiltonian can either be partitioned into subsets $Q_{ij}^{(N)} = \{ X_i X_j, Y_i Y_j, \{ Z_k\}_{k \ne i, j} \}_{i<j}$, or into $Q_{2}^{(N)} = \{  X_1 X_2,  X_1 X_3, ...,  X_2 X_3, ...,  X_{N-1}X_N \}$ and $Q_{3}^{(N)} =\{  Y_1 Y_2,  Y_1 Y_3, ..., Y_2 Y_3, ...,  Y_{N-1} Y_N \}$.
The set $Q_{2}^{(N)}$ yields a stabilizer energy equal to $-\frac{\vbar_x}{2(N-1)} \times \frac{N(N-1)}{2} = -\frac{N \vbar_x}{4}$, while 
$Q_{3}^{(N)}$ will always yield a larger energy~\footnote{With the optimal combination of signs we find stabilizer energy $-\frac{N\vbar_x}{4(N-1)}$ for even $N$ and $-\frac{\vbar_x}{4}$ for odd $N$.}.
A set of the former type $Q_{ij}^{(N)}$ yields energy $- \frac{(N-2)}{2} - \frac{\vbar_x}{(N-1)}$ which is also always larger when $\vbar_x >2$. Additionally, as mentioned above, those sets lead to parity-symmetry breaking which is not desired.
\\

In summary, for arbitrary value $N>2$, if $-\frac{N \bar{v}}{4} > -\frac{N}{2}$, i.e., $\vbar_x < 2 $, the stabilizer ground state is $\ket{\Psi_{s,1}}^{(N)} = \ket{1}^{\otimes N}$ with stabilizer energy $E_{s,1}= -N/2$, while if $\vbar_x > 2$ the stabilizer ground state $\ket{\Psi_{s,2}}^{(N)}$ is entangled with $E_{s,2}=-\frac{N \bar{v}}{4}$. For $N$ even (resp. odd), we can predict that this entangled state is an equal superposition of all states with even (resp. odd) number of spin down (1's).

Rigorously, to fully specify $\ket{\Psi_{s,2}}^{(N)}$, one needs a complete stabilizer group. To do this, $N-1$ independent generators can be extracted from the group $Q_{2}^{(N)} =\{ X_1 X_2, X_1 X_3, ... X_2 X_3, ..., X_{N-1}X_N \}$, for example: 
\begin{align}
    & g_1 = X_1 X_N  \; ,\nonumber \\
    & g_2 = X_2 X_N  \; ,\nonumber \\
    & ... \nonumber \\
    & g_{N-1} = X_{N-1} X_N \; ,
\end{align}
which can be completed by 
\begin{align}
& g_{N} = (-1)^N Z_1 Z_2...Z_N \; ,
\label{eq:gen_N}
\end{align}
(which satisfies permutation invariance) to form the $N$ generators of the complete stabilizer group $\mathcal{S}_2^{(N)} = \langle g_1, g_2, .., g_N$ of $\ket{\Psi_{s,2}}^{(N)}$.
\\
The preparation of this state can be inferred from generalizing expressions~\eqref{eq:state_prep_N2} and~\eqref{eq:state_prep_N3} as
\begin{align}
        \ket{\Psi_{s,2}}^{(N)} = 
        \begin{cases} 
        \prod_{i<j} \text{CX}_{ji} \ket{0} \otimes \ket{+}^{\otimes (N-1)} & \text{for even } N \; , \\ \\
         \prod_{i<j} \text{CX}_{ji} \ket{1} \otimes \ket{+}^{\otimes (N-1)} & \text{for odd } N  \nonumber \; .
        \end{cases}
\label{eq:stab_LMG_app} 
\end{align}
For example, for $N=4$, this gives
\begin{align}
    \ket{\Psi_{s,2}}^{(4)} 
    &= \frac{1}{2 \sqrt{2}} \left( \ket{0000} + \ket{0011} + \ket{0110} + \ket{1100} + \ket{0101} + \ket{1111} + \ket{1010} + \ket{1001} \right) \; .
\end{align}

The states $\ket{\Psi_s}^{(N)}$, are, as expected, equal superposition of configurations with even (resp. odd) spins down (1's) for $N$ even (resp. odd). They are also superpositions of the angular momentum eigenstates (Dicke state) $ \ket{J,M} \equiv \ket{N_+=J+M} $. 
For example, $\ket{\Psi_{s,2}}^{(2)}$ is a superposition of 
$\ket{N_+=0,2} = \ket{J=1, M= -1, +1}$, 
$\ket{\Psi_{s,2}}^{(3)}$ is a superposition of $\ket{N_+=0,2} = \ket{J=3/2, M= -3/2, +1/2}$ ($M=-1/2$ or $+3/2$ would break parity symmetry), and $\ket{\Psi_{s,2}}^{(4)}$ is a superposition of $\ket{N_+=0,2,4} = \ket{J=2, M= -2, 0, +2}$. 
\\
\\

The case $\chi \in (-1, 0)$ can be treated in exactly the same way as $\chi=-1$. 

Only in the case $\chi = +1$ would the states stabilized by
$\mathcal{S}_2^{(N)} = \langle X_1 X_N, X_2 X_N, ..., X_{N-1}X_N, (-1)^N Z_1 Z_2...Z_N \rangle $
and 
$\mathcal{S}_3^{(N)} = \langle Y_1 Y_N, Y_2 Y_N, ..., Y_{N-1}Y_N, (-1)^N Z_1 Z_2...Z_N \rangle $
be degenerate.

\subsection{Preparation of stabilizer ground states from graph states}

In this section we will omit superfluous super- and sub-scripts and refer to the $N$-spin entangled stabilizer ground state $\ket{\Psi_{s,2}}^{(N)}$ and corresponding stabilizer group $\mathcal{S}_2^{(N)}$,  simply as $\ket{\Psi_s}$ and $\mathcal{S}$, respectively.

In the case of the present LMG model, one has enough intuition to infer a way of preparing the entangled stabilizer ground state $\ket{\Psi_s}$, as written in Eq.~\eqref{eq:stab_LMG_app} above.
In the case of a more general Hamiltonian, however, the stabilizer state preparation requires a more systematic procedure. For example, it is known that any stabilizer state can be obtained via local Clifford operations acting on a graph state.

As detailed in, {\it e.g.}, Ref.~\cite{ed1d1be3cb1a475fb9d87d2369dfef15}, an $N$-qubit graph state $\ket{G}$ is associated with a (simple unweighted) graph $G = (V,E)$, where $V$ denotes the sets of vertices ($N$ qubits) and $E$ denotes the set of edges between these vertices.
The graph state $\ket{G}$ can be prepared as
\begin{align}
    \ket{G} = \left( \prod_{e \in E } \text{CZ}_e \right) \ket{+}^{\otimes N} \; ,
\end{align}
and is stabilized by operators of the form\footnote{This can be checked easily using $X_i \ket{+}_i = \ket{+}_i$ and $\text{CZ}_{ij} X_i = X_i Z_i \text{CZ}_{ij}$.}
\begin{align}
    g_i^{G} = X_i \bigotimes_{j \in n_i} Z_j \; ,
\end{align}
where $n_i$ denotes the set of qubits (vertices) connected to $i$ by an edge.
\\

In order to find the Clifford operations $U_C$ that transforms a graph state $\ket{G}$ into the stabilizer state $\ket{\Psi_s}$ as:
\begin{eqnarray}
    \ket{\Psi_s} = U_C \ket{G} \; ,
\label{eq:Cliff_graph_stab}
\end{eqnarray}
one can for example parametrize $U_C$ with a set of discrete angles (to ensure the Clifford-ness) and optimize the angles via energy minimization. Optimization procedures, in particular discrete ones, can however often be difficult to achieve in a robust way.

On the other hand, there exist an efficient procedure detailed in Ref.~\cite{PhysRevA.69.022316} to find the local Clifford operations that transform from a stabilizer state to a graph state, without resorting to an optimization. This procedure utilizes the stabilizer tableau formalism, i.e. the fact that a stabilizer state $\ket{\Psi_s}$ can be represented by a binary tableau, or generator matrix, 
\begin{equation}
    \bm{T}_s = \left( \bm{X} | \bm{Z} \right) \; ,
\end{equation}
where each line of $\bm{T}_s$ corresponds to a generator $g_i$ of the stabilizer group $\mathcal{S} = \langle g_1, g_2, ..., g_N \rangle$ of $\ket{\Psi_s}$, and the matrix elements  $\bm{X}_{ij} = 1$ if $g_i$ acts with X on qubit $j$, and similarly $\bm{Z}_{ij} = 1$ if $g_i$ acts with Z on qubit $j$. 
For example, the binary tableau representing $X_1 Y_2 Z_3$ for a three-qubit state is $\left( 1 1 0 | 0 1 1\right)$. Note that the sign of each generator is ignored.

The generator matrix associated with a graph state $\ket{G}$ has the following particular structure
\begin{equation}
    \bm{T}_G = \left( \bm{\mathds{1}} | \bm{\Gamma} \right) \; ,
\end{equation}
where $\Gamma$ is the adjacency matrix of the graph $G=(V,E)$, {\it i.e.}
\begin{align}
        \Gamma_{ij} = 
        \begin{cases} 
        1 & \text{if } (i,j) \in E \; , \\ 
        0 & \text{otherwise}  \nonumber \; .
        \end{cases} 
\end{align}
Since we are dealing with simple (unweighted) graphs which have no loops, $\Gamma_{ii}=0$.
\\
Ref.~\cite{PhysRevA.69.022316} describes the (efficient) procedure for transforming the generator matrix of a stabilizer state into one a graph state:
\begin{align}
    \bm{T}_s = \left( \bm{X} | \bm{Z} \right) \rightarrow \bm{T}_G = \left( \bm{\mathds{1}} | \bm{\Gamma} \right)  \; .
\end{align}
This directly provides the desired transformation $U_C$ in Eq.~\eqref{eq:Cliff_graph_stab}.
\\

In the present study of the LMG model, the generator matrix representing the stabilizer group $\mathcal{S}= \langle X_1 X_N, X_2 X_N, ..., X_{N-1} X_N, (-1)^N Z_1 Z_2...Z_N \rangle$ associated with the entangled stabilizer state $\ket{\Psi_s}$ is then given by
\begin{equation}
\bm{T}_s =
\left(
\begin{array}{cccccc|cccccc}
    1 & 0 & 0 & ... & 0 & 1 & 0 & 0 & 0 & ... & 0 & 0  \\ 
    0 & 1 & 0 & ... & 0 & 1 & 0 & 0 & 0 & ... & 0 & 0  \\
    0 & 0 & 1 & ... & 0 & 1 & 0 & 0 & 0 & ... & 0 & 0  \\ 
      &   &   & ... &   &   &   &   &   &...  &   &   \\
    0 & 0 & 0 & ... & 1 & 1 & 0 & 0 & 0 & ... & 0 & 0  \\ 
    0 & 0 & 0 & ... & 0 & 0 & 1 & 1 & 1 & ... & 1 & 1  \\ 
\end{array}
\right) \; .
\end{equation}

The procedure in Ref.~\cite{PhysRevA.69.022316} then simply amounts to applying a Hadamard gate on the last qubit. The effect of Hadamard conjugation on the binary tableau is to exchange the last columns of the matrices $\bm{X}$ and $\bm{Z}$, yielding
\begin{equation}
\bm{T}_G =
\left(
\begin{array}{cccccc|cccccc}
    1 & 0 & 0 & ... & 0 & 0 & 0 & 0 & 0 & ... & 0 & 1  \\ 
    0 & 1 & 0 & ... & 0 & 0 & 0 & 0 & 0 & ... & 0 & 1  \\
    0 & 0 & 1 & ... & 0 & 0 & 0 & 0 & 0 & ... & 0 & 1  \\ 
      &   &   & ... &   &   &   &   &   &...  &   &   \\
    0 & 0 & 0 & ... & 1 & 0 & 0 & 0 & 0 & ... & 0 & 1  \\ 
    0 & 0 & 0 & ... & 0 & 1 & 1 & 1 & 1 & ... & 1 & 0  \\ 
\end{array}
\right) 
= \left( \bm{\mathds{1}} | \bm{\Gamma} \right) \; ,
\label{eq:stab_matrix_G}
\end{equation}
which describes a graph state with stabilizer group $\mathcal{S}_G = \{ X_1 Z_N, X_2 Z_N, ..., X_{N_1} Z_N, Z_1 ... Z_{N-1} X_N \}$ which is indeed the conjugate of $\mathcal{S}$ with $H_N$, for even values of $N$. 
In the case when $N$ is odd, one needs to apply an extra X gate to one of the qubits. This has the effect of conjugating each generator with X and thus provides the necessary minus sign for the last generator $- Z_1 Z_2...Z_N$ (see Eq.~\eqref{eq:gen_N}). 
\\
In summary, if we denote by $g_i^{(G)}$ the generators in $\mathcal{S}_G$ and $g_i$ the generators in $\mathcal{S}$, we have
\begin{align}
    g_i^{(G)} &= U_{C}\,   g_i \,  U^\dagger_{C} \; ,
\end{align}
or equivalently,
\begin{align}
    \ket{G} &= U_{C} \, \ket{\Psi_s}^{(N)} \; ,
\end{align}
where, for example,
\begin{align}
    U_C = (X_1)^{N \, \text{mod}\, 2} \, H_N \; .
\end{align}

The generator matrix in Eq.~\eqref{eq:stab_matrix_G} corresponds to an open graph with edges between vertex (qubit) $N$ and all vertices $i<N$. This is shown in panel a) of Fig.~\ref{fig:graph_N8} for the case $N=8$.
Note that graph $G$ is the complementary graph to the all-to-all connected graph $G^c$, as shown in panel b) of Fig.~\ref{fig:graph_N8}. The adjacency matrix of $G$ and $G^c$ are related by~\cite{PhysRevA.69.022316}: $\bm{\Gamma}^c = \bm{\Gamma} + \bar{\mathds{1}}$ where 
\begin{align}
        \bar{\mathds{1}}_{ij} =  
        \begin{cases} 
        1 & \text{if } i \ne j \; , \\ 
        0 & \text{if } i = j  \nonumber \; .
        \end{cases} 
\end{align}

In summary, an alternative way of preparing the entangled stabilizer ground state of the LMG model is given by acting with a single Hadamard gate (and X if $N$ is odd) on the open graph state as
\begin{align}
     \ket{\Psi_s} & = \hat{U}_G^{-1} \, \ket{G} \; , \\
                        & = (X_1)^{N \, \text{mod}\, 2} \, H_N \, \prod_{i=1}^{N-1} \text{CZ}_{i, N} \ket{+}^{\otimes N} \; .
\label{eq:stab_prep_wgraph}
\end{align}
Unlike Eq.~\eqref{eq:stab_LMG_app}, the preparation given in Eq.~\eqref{eq:stab_prep_wgraph} has been obtained from following the procedure in Ref.~\cite{PhysRevA.69.022316} which provides a systematic and rigorous "recipe", generalizable to the case of an arbitrary Hamiltonian.

%%%%%%%%%%%%%%%%%%%%%%%
\section{ Results for Stabilizer Ground States and Beyond }
\label{sec:app_results_varQITP}
%%%%%%%%%%%%%%%%%%%%%%%

In this appendix we provide additional results obtained for systems ranging from $N=2$ to $30$ spins. We focus on the case $\chi=-1$, as the results do not significantly change for $\chi \in [-1, 0)$.

\begin{figure}[h]
     \centering
         \includegraphics[width=\textwidth]{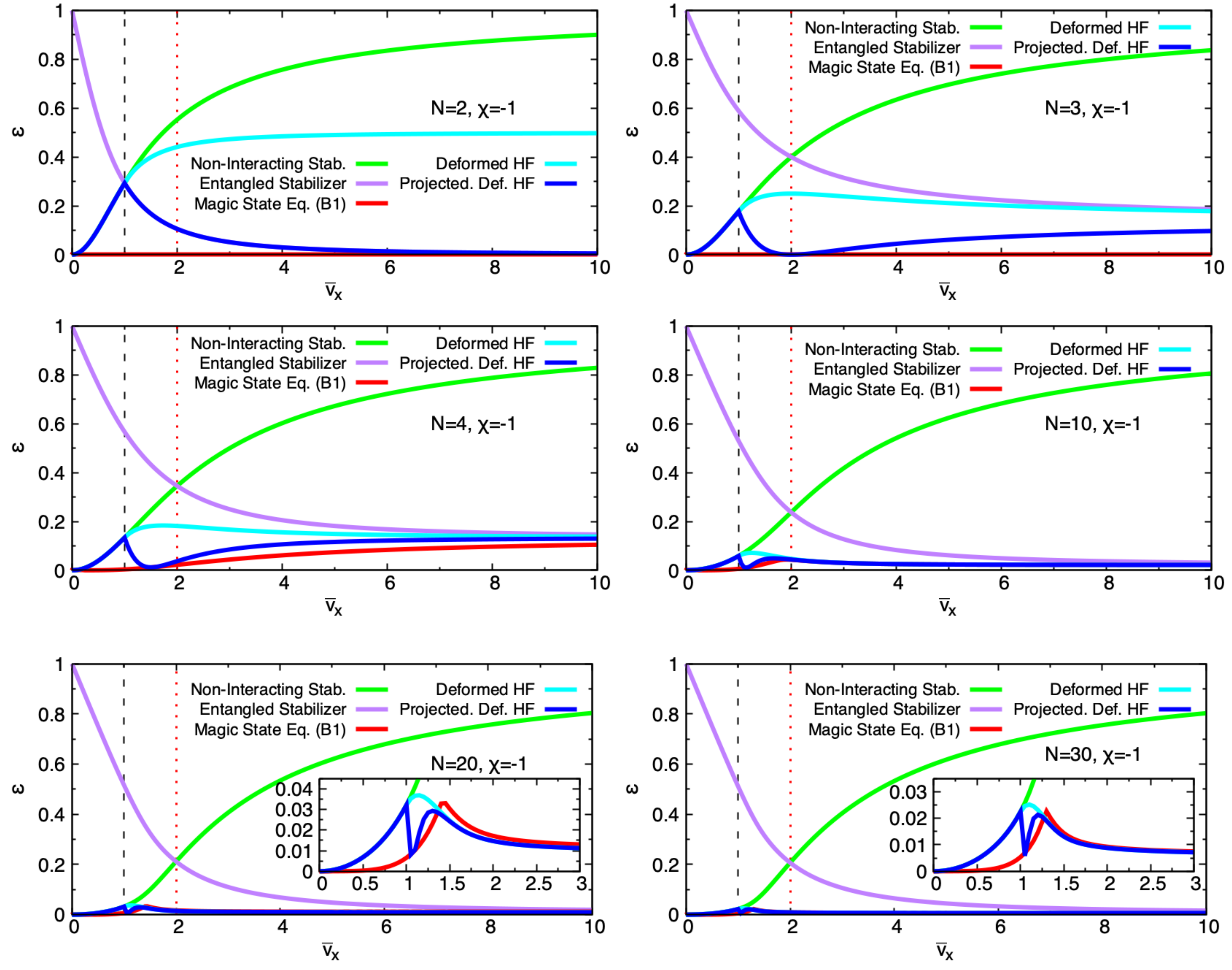}
    \caption{Relative energy difference between the exact ground-state energy and several approximations, for systems with $N=2, \, 3, \, 4, \, 10, \, 20, \, 30$ and $\chi= -1$. 
    The unentangled and entangled stabilizer states, $\ket{\Psi_{s,1}}^{(N)}$ and $\ket{\Psi_{s,2}}^{(N)}$, are shown in green and purple curves, respectively. The state obtained via Eq.~\eqref{eq:var_Jz_app} is shown with red curves. For comparison, the deformed HF without and with projection are shown with cyan and blue curves, respectively. }
    \label{fig:energy_diff_Jz}
\end{figure}

\begin{figure}[h]
     \centering
         \includegraphics[width=\textwidth]{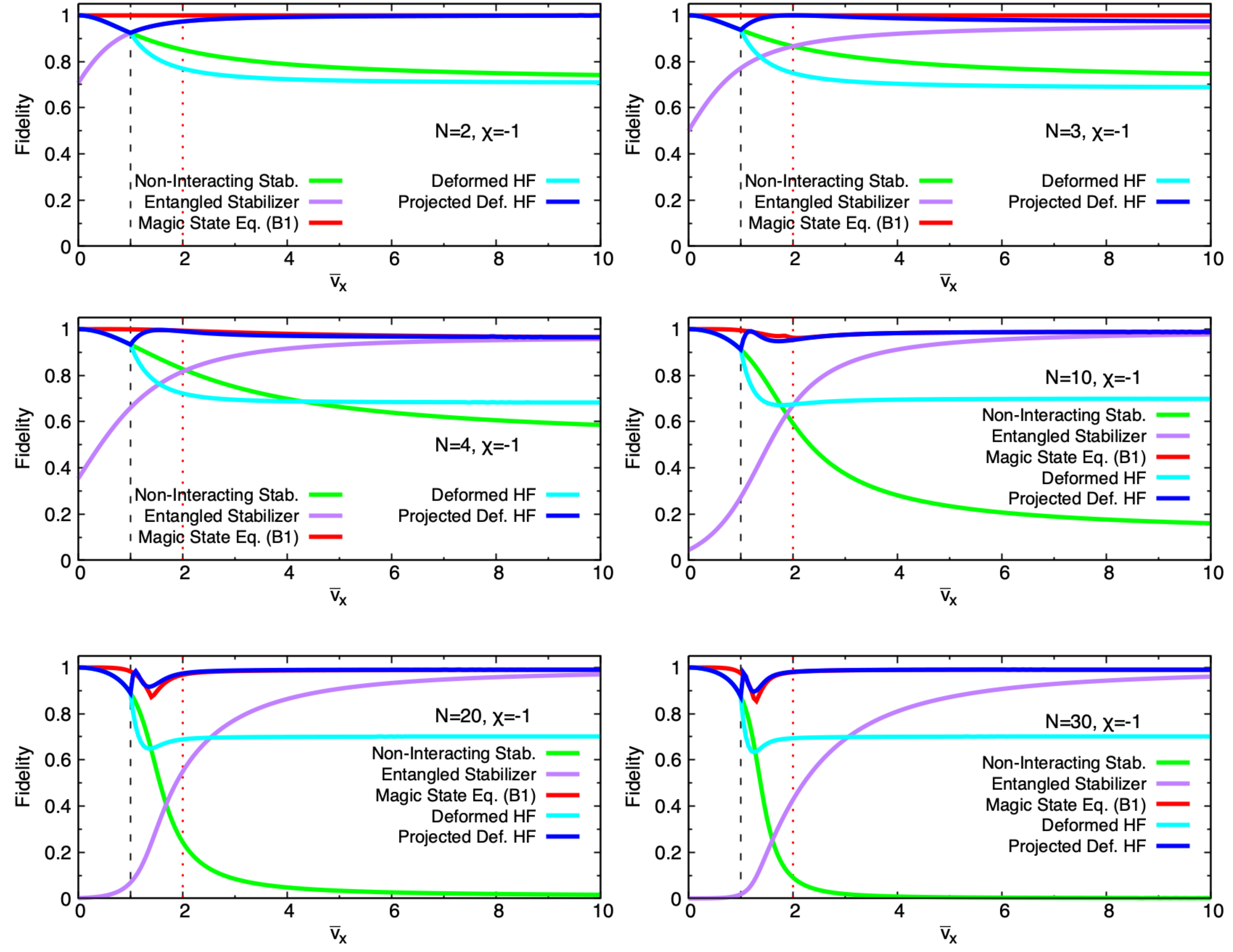}
         \caption{Same as Fig.~\ref{fig:energy_diff_Jz} for the fidelity of approximate states to the exact ground-state. }
         \label{fig:fidelity_Jz}
\end{figure}

Figs.~\ref{fig:energy_diff_Jz} and \ref{fig:fidelity_Jz} show the relative energy difference to the exact ground state $\varepsilon =  |(E_{app} - E_{ex}) / E_{ex}|$ and fidelity $|\langle  \Phi_{ex} | \Psi_{app} \rangle|$, respectively, for quantum states $\ket{\Psi_{app}}$ obtained at different levels of approximation.
In particular, results obtained for the non-interacting ground state (unentangled stabilizer state) $\ket{\Psi_{s,1}}^{(N)} = \ket{1}^{\otimes N}$ are shown with a light green curve, while those obtained for the entangled stabilizer state  $\ket{\Psi_{s,2}}^{(N)}$ are shown in purple.
The magic state obtained via
\begin{equation}
  \ket{\Phi(\theta)}^{(N)} = \text{e}^{- i \theta \hat{J}_z} \ket{\Psi_{s,2}}^{(N)} \; ,
  \label{eq:var_Jz_app}
\end{equation}
as discussed in Sec.~\ref{sec:Var_QITP}, is shown with a red curve.
Finally, for comparison, the deformed Hartree-Fock (HF) without and with projection onto a good parity are shown with a cyan and blue curve, respectively.

The exact solution and entangled stabilizer $\ket{\Psi_{s,2}}^{(N)}$ for systems with $N=2,3$ contain only two Dicke states, and thus differ by only one relative amplitude, which can be adjusted with a single angle. Thus Eq.~\eqref{eq:var_Jz_app} recovers the exact ground state energy and wave function for these systems.
This is true even in the region below the stabilizer ground state transition point ($\vbar_x < 1$ for $N=2$ and $\vbar_x < 2$ for $N=3$). 

Larger values of $N$, the magic states prepared via Eq.~\eqref{eq:var_Jz_app} show small deviations with the exact solution. The relative energy tends to improve and converge towards $\simeq 0$ for large $N>4$ values, while the discrepancy of the wave functions increases around the stabilizer-ground-state transition point around $\vbar_x \simeq 2$, where the magic of the exact state is maximum.  
We have observed the same trends, to a greater extent, for the stabilizer ground states themselves in Fig.~\ref{fig:Stabil_gs_Ns} of the main text. 

We note that such discrepancies can be systematically improved with higher-order operators, starting with the tensor operator $J_z^2$, which brings both entanglement and magic. In that case, the variational state takes the form
\begin{equation}
    \ket{\Psi (\theta_1, \theta_2)} = \text{e}^{-i \theta_2 J_z^2} \text{e}^{-i \theta_1 J_z} \ket{\Psi_{s,2}} \; .
\end{equation}

As noted in section \ref{sec:Stabilizer_gs_LMG}, we see that the fidelity crossing point decreases as $N$ increases. In fact, in the large $N$ limit, we find that this crossing point converges towards $\vbar_x =1$, which coincides with the phase transition, and that the value of the fidelity goes to zero at this point. Since the interaction strength is proportional to $\vbar_x/(N-1)$, the region where the stabilizer state becomes close to the ground state also becomes shifted towards larger values of $\vbar_x$. For fixed value of $\vbar_x /(N-1) \gtrsim 1$, the fidelity of the stabilizer ground state is roughly constant with $N$ and $ \simeq 0.98 - 0.99$.

%%%%%%%%%%%%%%%%%%%%%%%
\section{ QITP and variational variant }
\label{sec:app_QITP}
%%%%%%%%%%%%%%%%%%%%%%%
The Quantum Imaginary Time Propagation (QITP) developed in Ref.~\cite{Turro:2021vbk} is an algorithm implementing imaginary-time evolution (ITE) using quantum circuits.
Such technique is inspired by classical ITE methods, which, in their original form, are used to find the ground state of a physical system, by evolving an initial state $\ket{\eta} \equiv \ket{\eta(0)}$ (with non-zero overlap with the ground state) in imaginary time $\tau = i t$
\begin{equation}
    \text{e}^{-(\hat H -  \bar{E}_0) \tau} \ket{\eta} = \sum_n \text{e}^{-(E_n - \bar{E}_0) \tau} \langle \Phi_{n} | \eta \rangle  \ket{\Phi_{n}} \; ,
\label{eq:QITE}
\end{equation}
where $\ket{\Phi_n}$ denote eigenstates of $\hat H$, in particular $\ket{\Phi_0} \equiv \ket{\Phi_{ex}}$ is the exact ground state, and $\bar{E}_0$ is an upper bound close the exact ground state energy.
After a long time $\tau$, the excited states components are suppressed and the system reaches its ground state 
\begin{equation}
    \lim_{\tau \rightarrow \infty} \text{e}^{-(\hat H -  \bar{E}_0) \tau} \ket{\eta} = \langle \Phi_{0} | \eta \rangle  \ket{\Phi_{0}} \; .
\end{equation}

Since the operator in Eq.~\eqref{eq:QITE} is not unitary, it cannot be straightforwardly implemented on a quantum computer. 
 The QITP algorithm circumvents this issue by embedding the system of interest into a larger one by adding an ancilla qubit (initialized to state $\ket{0}$). One then acts on the full enlarged system with a unitary operator $\hat U$ chosen such that the action of $\hat U$ yields Eq.~\eqref{eq:QITE} on the system of interest, after measurement of the ancilla qubit. 
 QITP is described in detail in Ref.~\cite{Turro:2021vbk}, and we simply remind the main steps below.
Precisely, 
\begin{eqnarray}
    \hat U (\tau) = 
    \begin{pmatrix}
        \hat Q (\tau) & \hat A (\tau) \\
        \hat A (\tau) & - \hat Q (\tau)
    \end{pmatrix}
\label{eq:U_QITP}
\end{eqnarray}
where 
\begin{align}
    \hat A (\tau) &= \left( \mathds{1} + \text{e}^{-2(\hat H -  \bar{E}_0) \tau} \right)^{-1/2} \; , \\
    \hat Q (\tau) &= \hat A (\tau) \, \text{e}^{-(\hat H -  \bar{E}_0) \tau} = \left( \mathds{1} + \text{e}^{2(\hat H -  \bar{E}_0) \tau}  \right)^{-1/2} \; .
\end{align}
Measuring the ancilla qubit after acting with $\hat U$ on the full system $\ket{0} \otimes \ket{\eta}$, and keeping only the samples that have collapsed to state $\ket{0}$, the state of the system of interest collapses to~\cite{Turro:2021vbk}
\begin{align}
    \ket{ \eta(\tau) } &=  \frac{\hat Q (\tau) \ket{\eta}}{||\hat Q (\tau) \ket{\eta}||} \; , \text{ with }
     \hat Q (\tau) \ket{\eta} \xrightarrow[\tau \rightarrow \infty]{} \frac{\langle \Phi_{0} | \eta \rangle}{\sqrt{2}}  \ket{\Phi_{0}} \; ,
\label{eq:QITE_final_state}
\end{align}
which is the desired ground state. The success of the QITP procedure depends on the squared overlap of the initial state with the exact ground state $| \langle \Phi_{0} | \eta \rangle |^2$.
Choosing the unentangled and entangled stabilizer states, $\ket{\Psi_{s,1}}^{(N)}$ and $\ket{\Psi_{s,2}}^{(N)}$, as initial states, the QITP procedure yields the results shown in Fig.~\ref{fig:full_QITP} of the main text. 
Note that in a real quantum circuit one typically has to resort to Trotterization of the evolution operator. Implementing and investigating the impact of such Trotterization is outside the scope of the present study.
\\

\begin{figure}[h]
     \centering
         \includegraphics[width=0.75\columnwidth]{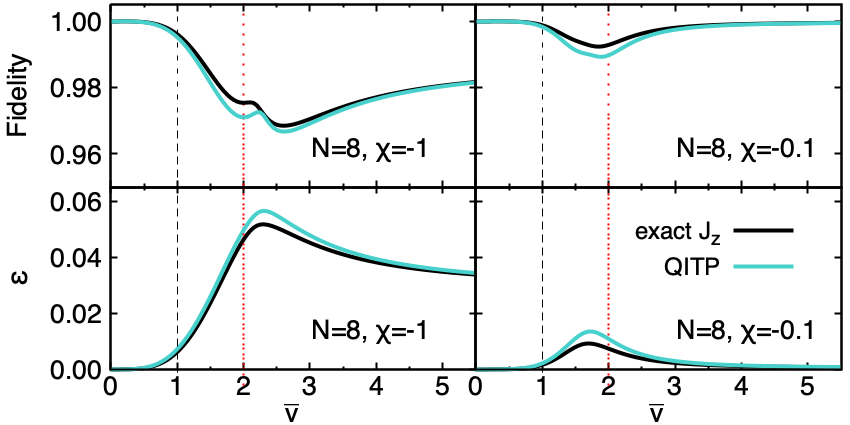}
         \caption{Comparison of the relative energy difference and fidelity obtained with implementing Eq.~\eqref{eq:var_QITP_Jz} exactly (black curve) and using the variational QITP operator of Eq.~\eqref{eq:state_varQITP} (teal curve) 
         for a system with $N=8$ spins, and parameter values  $\chi=-1$ (left) and $\chi=-0.1$ (right). }
         \label{fig:comp_varJz_QITP}
\end{figure}

Further, the QITP procedure described above can also be utilized and adapted in order to implement Eq.~\eqref{eq:var_QITP_Jz} on a quantum circuit.
Precisely the operator $\text{exp}(- \hat{J}_z \theta)$ can be implemented via Eqs.~\eqref{eq:U_QITP}-\eqref{eq:QITE_final_state} by making the substitutions
\begin{align}
    \hat H \rightarrow \hat{J}_z \; , \; \; \text{and, } \; \;
    \tau \rightarrow \theta \; ,
\end{align}
so that the system is evolved with the $\hat{J}_z$ operator instead of the full Hamiltonian, and the angle $\theta$ plays the role of the imaginary time $\tau$. 
The state of the system after measurement of the ancilla qubit is then given by
\begin{align}
    \ket{\Phi (\theta)} = \hat Q (\theta) \ket{\Psi_{s,2}}^{(N)} \; .
    \label{eq:state_varQITP}
\end{align}
The difference with full QITP is that the angle $\theta$ does not go to infinity. Instead there is an optimal value for $\theta$ which provides a minimal energy, and thus can be determined via an optimization procedure. 
As mentioned in the main text, in practice, the optimal value of $\theta$ has been obtained by manually scanning the energies obtained for a range of discrete values of $\theta$, followed by a further refinement using a Scipy optimizer~\cite{2020SciPy-NMeth}. In particular, we have tested both Cobyla and L-BFGS-B which have provided the same results. 
Fig.~\ref{fig:comp_varJz_QITP} shows a comparison of the results obtained with implementing Eq.~\eqref{eq:var_QITP_Jz} exactly and using the QITP algorithm as in Eq.~\eqref{eq:state_varQITP}, for the two parameter values  $\chi=-1$ and $\chi=-0.1$. Except for small discrepancies occur around $\vbar_x =2$ (the most magic region), the QITP algorithm reproduces the exact evolution to a large extent.
%

%\clearpage
%%%%%%%%%%%%%%%%%%%%%%%
\section{ADAPT-VQE on a Stabilizer Ground State}
\label{sec:app_adapt}
%%%%%%%%%%%%%%%%%%%%%%%

ADAPT-VQE has been originally proposed in Ref.~\cite{Grimsley:2018wnd}. This classical-quantum algorithm is based on a wave fucntion ansatz of the form
\begin{align}
    \ket{\Phi (\theta_1,...,\theta_L)} = \prod_{l = 1}^L \text{e}^{i \theta_l \hat{T}_l} \ket{\phi} \; ,
\end{align}
where $\ket{\phi}$ is a chosen reference state, and the $\hat{T}_l$'s are Hermitian excitation operators, chosen from a pre-defined pool. At each layer $l$, an excitation operator is selected based on its energy gradient 
\begin{equation}
    \frac{\partial E}{\partial \theta_l}_{| \theta_l = 0} = - i   \langle \Phi_{l -1 } | \left[ \hat T_{l}, \hat H \right] |\Phi_{l -1 }\rangle   \; ,
    \label{eq:adapt_comm}
\end{equation}
where $ \ket{\Phi_{l -1 }} \equiv \ket{\Phi (\theta_1,...,\theta_{l-1})} $.
At each layer, the values of the angles $\theta_l$ are optimized via a VQE algorithm~\cite{Peruzzo:2013bzg}.
\\

In this section we explore the use of ADAPT-VQE when the reference state $\ket{\phi}$ is taken to be the stabilizer ground state. We restrict ourselves to the region $\vbar_x >2$, for which the stabilizer ground state is entangled, specifically, $\ket{\phi} = \ket{\Psi_{s,2}}$, given in Eqs.~\eqref{eq:stab_LMG} and \eqref{eq:stab_LMG_graph}.
We limit the operator pool to operators preserving the relevant symmetries (parity and permutation symmetries), and the real character of the wave function. These are 
\begin{equation}
    \hat{T}_{ij}^\pm = X_i Y_j \pm Y_i X_j \; .
\label{eq:ops-adapt}
\end{equation}
We note, however, that allowing for a larger pool, including operators such as $\frac{1}{2} (XX \pm YY)$ or symmetry-braking ones such as $XZ$, $YZ$, does not change the results as the associated gradients are always found to be smaller than those of operators in Eq.~\eqref{eq:ops-adapt}, and thus are never selected during the procedure~\footnote{Since the energy gradient is given by the expectation value of the commutator of the Hamiltonian and the excitation operator (Eq.~\eqref{eq:adapt_comm}), ADAPT-VQE can never bring towards a symmetry-broken solution if starting from a symmetry-preserving one.}.

\begin{figure}
     \centering
         \includegraphics[width=0.9\columnwidth]{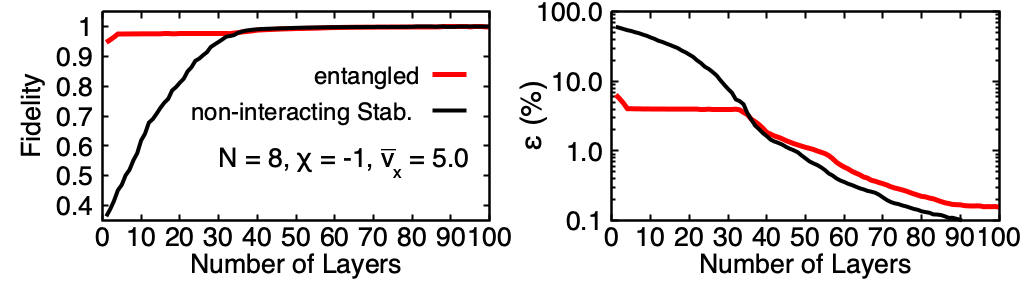}
         \caption{ADAPT-VQE convergence for a system with $N=8$ spins, $\chi=-1$, and $ \vbar_x = 5$. The left and right panel show the fidelity and relative energy error of the ADAPT-VQE wave function, receptively. The black curve shows the results obtained with the unentangled reference state $\ket{\phi} = \ket{\Psi_{s,1}}$, while the red curve shows the results for $\ket{\phi} = \ket{\Psi_{s,2}}$. }
         \label{fig:adapt-vqe}
\end{figure}

As mentioned in the main text, we find that such implementation of ADAPT-VQE, on top of a collectively entangled stabilizer state, is not adequate, as the layers somewhat destroy the collectivity of the state. This is illustrated in Fig.~\ref{fig:adapt-vqe}, which shows, as an example, results obtained for a system with $N=8$ spins, $\chi=-1$ in the region of large deformation at $ \vbar_x = 5$.
The results obtained with $\ket{\phi} = \ket{\Psi_{s,2}}$ are shown with red curves. For comparison we also show with black curves the results obtained  with $\ket{\phi} = \ket{\Psi_{s,1}} = \ket{1}^{\otimes 8}$.
Although $\ket{\Psi_{s,2}}$ is closer to the exact solution, the gradients of the various operators are found to be of similar and very small magnitude for a large number of layers, leading to the observed plateau in both the total energy and wave function. It is only after $\simeq 35$ layers, where both solutions meet, that the gradients increase and convergence towards the exact ground state begins. Typically we find that convergence then takes place at a similar rate as with the unentangled reference state (slightly slower or faster, depending on the parameter values).

In any case, it seen that the number of layers of order 100 required to reach $0.1\%$ accuracy in the energy is too large to be implemented on real quantum devices. 
Thus, applying APADT-VQE with one- and two-spin excitation operators appears to be inadequate for the case of collective systems with all-to-all connectivity. If possible, one would ideally use collective excitation operators~\cite{Farrell:2023fgd}. In general, this is however not straightforward to implement due to the non-commutativity of the individual terms in such operators.

Whether such slow convergence is specific to deformation remains to be seen. We note that Ref.~\cite{Zhang:2024uxp} applied ADAPT-VQE with various operator pools to the like-particle and neutron-proton pairing problem in reasonable numbers of iterations (except in some degenerate cases).

\end{document}